\documentclass[10.5pt,letterpaper]{article}
\pdfoutput=1 

\usepackage[utf8]{inputenc} 

\usepackage[top=2.5cm, bottom=3cm, left=3.5cm, right=3.5cm,
           heightrounded,marginparwidth=1.5cm, marginparsep=1cm]{geometry} 

\usepackage{changepage} 
\usepackage{etoolbox}

\usepackage[shortlabels]{enumitem} 

\usepackage[square,numbers,merge,comma,sort&compress]{natbib} 
\makeatletter
\def\NAT@spacechar{\,}  
\makeatother

\usepackage{amsmath,amssymb,amsfonts,amsthm}
\usepackage{mathtools} 
\usepackage{old-arrows}
\usepackage[scale=0.7]{miama} 
\usepackage{slashed,cancel}
\usepackage{comment}   
\usepackage{relsize}   
\usepackage{setspace}  
\usepackage{moresize}  
\usepackage{epsfig}
\usepackage{latexsym}
\usepackage{mathrsfs,calligra,aurical} 
\usepackage{calc}
\usepackage{float}
\usepackage{appendix}
\usepackage{xargs}
\usepackage{extarrows}
\usepackage{empheq}

\usepackage{graphicx} 

\usepackage[labelsep=colon]{caption}  
\usepackage[labelsep=colon,aboveskip=10pt,belowskip=10pt]{subcaption}
\DeclareCaptionSubType * [roman]{table}
\usepackage{array,multirow,makecell,booktabs}  
\newcolumntype{X}[2]{>{\centering\arraybackslash$}#1{#2\linewidth}<{$}}
\newcolumntype{R}[1]{>{\raggedleft\arraybackslash$}m{#1\linewidth}<{$}}
\newcolumntype{L}[1]{>{\raggedright\arraybackslash}m{#1\linewidth}}

\usepackage{titlesec}

\titleformat{\section}{\normalfont\large\bfseries}{\thesection}{1em}{}
\titleformat{\subsection}{\normalfont\normalsize\bfseries}{\thesubsection}{0.75em}{}
\titleformat{\subsubsection}{\normalfont\normalsize\bfseries}{\thesubsubsection}{0.75em}{}

\titlespacing*{\section}{0pt}%
                {4ex plus 1ex minus .5ex}{1.75ex plus .25ex minus .25ex}
\titlespacing*{\subsection}{0pt}%
                {3.5ex plus 1ex minus .5ex}{1.25ex plus .2ex minus .2ex}
\titlespacing*{\subsubsection}{0pt}%
                {2.5ex plus 0.75ex minus .2ex}{0.75ex plus .15ex minus .15ex}
\titlespacing*{\paragraph}{0pt}%
                {1.85ex plus 0.5ex minus .15ex}{1em}

\usepackage{titletoc}

\addtocontents{toc}{\addvspace{-0.75em}}  

\titlecontents{section}
  [1.25em] {\addvspace{0.7em plus 0pt}\small}
  {\thecontentslabel\hspace{0.75em}}{}
  {\hspace{0.5em}\titlerule*[0.5em]{.}\contentspage}
  [\addvspace{0.0em plus 0pt}]

\titlecontents{subsection}
  [2.75em] {\addvspace{0.075em plus0pt}\fns}
  {\thecontentslabel\hspace{0.75em}}{\thecontentslabel\hspace{0.75em}}
  {\hspace{0.5em}\titlerule*[0.5em]{.}\small\contentspage}
  [\addvspace{0.075em plus 0pt}]

\usepackage{environ}
\makeatletter
\NewEnviron{subalign}[1]{
\begin{subequations}\label{#1}
\begin{align} \BODY \end{align}
\end{subequations}      }
\makeatother
\makeatletter
{\begingroup%
\setlength{\abovedisplayskip}{10pt plus 4pt minus 9pt}%
\setlength{\abovedisplayshortskip}{0pt plus 2pt minus 2pt}%
\setlength{\belowdisplayskip}{12pt plus 3pt minus 9pt}%
\setlength{\belowdisplayshortskip}{7pt plus 3pt minus 4pt}%
\begin{subequations}%
}%
{\end{subequations}\ignorespacesafterend%
\endgroup}%
\makeatother

\makeatletter
{\begingroup%
\setlength{\abovedisplayskip}{{#1}pt plus 2pt minus 9pt}%
\setlength{\abovedisplayshortskip}{0pt plus 0pt minus 2pt}%
\setlength{\belowdisplayskip}{{#2}pt plus 3pt minus 9pt}%
\setlength{\belowdisplayshortskip}{7pt plus 3pt minus 4pt}%
\begin{subequations}%
}%
{\end{subequations}\ignorespacesafterend%
\endgroup}%
\makeatother

\makeatletter
{\begingroup%
\setlength{\abovedisplayskip}{{#1}pt plus 3pt minus 9pt}%
\setlength{\abovedisplayshortskip}{0pt plus 3pt}%
\setlength{\belowdisplayskip}{{#2}pt plus 3pt minus 9pt}%
\setlength{\belowdisplayshortskip}{7pt plus 3pt minus 4pt}%
\begin{equation}%
}%
{\end{equation}\ignorespacesafterend%
\endgroup}%
\makeatother

\makeatletter
{%
\begin{equation}%
\begin{split}{#1}%
}%
{\end{split}%
\end{equation}\ignorespacesafterend%
}%
\makeatother

\usepackage[svgnames]{xcolor}
\definecolor{Green}{rgb}{0.05, 0.45, 0.25}
\definecolor{dogwoodrose}{rgb}{0.8, 0.1, 0.55}
\definecolor{RRed}{rgb}{0.7, 0.1, 0.525}

\usepackage{bm}  
\usepackage{dsfont}  

\DeclareMathAlphabet{\mathpzc}{OT1}{pzc}{m}{it}
\DeclareMathAlphabet{\mathcal}{OMS}{cmsy}{m}{n}
\DeclareSymbolFontAlphabet{\Scr}{rsfs}
\DeclareMathAlphabet{\mathbold}{U}{BOONDOX-ds}{m}{n}
\SetMathAlphabet{\mathbold}{bold}{U}{BOONDOX-ds}{b}{n}
\DeclareMathAlphabet{\mathcalboondox}{U}{BOONDOX-calo}{m}{n}
\SetMathAlphabet{\mathcalboondox}{bold}{U}{BOONDOX-calo}{b}{n}
\DeclareMathAlphabet{\mathbcalboondox}{U}{BOONDOX-calo}{b}{n}

\newcommand\eqlinkcol{RRed}



\usepackage[breaklinks=true,backref=page]{hyperref}
\hypersetup{
    bookmarks=true,         
    pdfmenubar=true,        
    pdffitwindow=false,     
    pdfpagemode={UseNone},
    pdfstartview={FitH},
    pdfauthor={},     
    pdfsubject={},   
    pdfcreator={},   
    pdfproducer={},  
    colorlinks=true,
    bookmarks=true,
    bookmarksnumbered=true,
    plainpages,
    a4paper,
    linktoc=page,
    citecolor=blue,
    filecolor=black,
    linkcolor=\eqlinkcol,
    urlcolor=Green,
}
\renewcommand*{\backref}[1]{}
\renewcommand*{\backrefalt}[4]{%
\ifcase #1 %
\relax
\or
~{\small [\textsc{p.~\fns{\!#2}}]}
\else
~{\small [\textsc{p.~\fns{\!#2}}]}%
\fi}

\usepackage{footnotebackref}
\usepackage{hypernat} 

\usepackage{cleveref} 
\crefname{equation}{}{}

\def\+{\:+\:}
\def\-{\:-\:}
\def\={\:=\:}
\def\'{``}
\def\*{{}^*}
\newcommand\Hodge{{}^*}
\newcommand\fns{\footnotesize}
\newcommand\qqquad{\quad\quad\quad}
\newcommand\qqqquad{\quad\quad\quad\quad}

\newcommand\Real{\operatorname{Re}}
\newcommand\Img{\operatorname{Im}}

\newcommand{\ml}{\mathlarger}

\newcommand{\dd}{\partial}
\providecommand{\abs}[1]{\lvert#1\rvert}

\newcommandx{\tcr}[1]{\textcolor{Crimson}{#1}}

\newcommand\eps{\epsilon}
\newcommand\veps{\varepsilon}
\newcommand\w{\omega}
\newcommand\Id{\mathds{1}}
\newcommand\Zero{\mathbold{0}}
\newcommand\N{\mathcal{N}}

\newcommand\nv{n_\text{v}}
\newcommand\ns{n_\text{s}}
\newcommand\II{\mathcal{I}}
\newcommand\RR{\mathcal{R}}
\newcommand\Ms{\mathscr{M}}

\newcommand\D{\mathcal{D}}

\newcommand\Lagr{\mathscr{L}}

\newcommand\M{\mathcal{M}}
\newcommand\FF{\mathbb{F}}
\newcommand\Cc{\mathbb{C}}

\newcommand\U{\mathrm{U}}
\newcommand\SL{\mathrm{SL}}

\newcommand\SO{\mathrm{SO}}

\newcommand\Rsvst{\mathscr{R}_\text{v*}}
\newcommand\VBH{V_\textsc{bh}}

\newcommand\Zch{\mathscr{Z}}

\newcommand\V{\mathcal{V}}
\newcommand\W{\mathcal{W}}
\newcommand\F{\mathcal{F}}
\newcommand\Fbo{\mathcalboondox{F}}
\newcommand\X{\mathcal{X}}
\newcommand\Ucal{\mathcal{U}}
\newcommand\zb{{\bar{z}}}
\newcommand\ib{{\bar{\imath}}}
\newcommand\jb{{\bar{\jmath}}}

\newcommandx{\sh}[1][1=\alpha,usedefault]{\sinh\left(#1\right)}
\newcommandx{\ch}[1][1=\alpha,usedefault]{\cosh\left(#1\right)}
\newcommandx{\sech}[1][1=\alpha,usedefault]{\operatorname{sech}\left(#1\right)}
\newcommandx{\cosech}[1][1=\alpha,usedefault]{\operatorname{cosech}\left(#1\right)}
\newcommandx{\tts}[1]{\text{\textsmaller{#1}}}
\newcommandx{\dm}[1][1=\mu,usedefault]{\partial_{#1}}
\newcommandx{\dmup}[1][1=\mu,usedefault]{\partial^{#1}}
\newcommandx{\subm}[2][1=p,2=A,usedefault]{{#1}_{\!\mathsmaller{#2}}}
\newcommandx{\subt}[2][1=p,2=A,usedefault]{{#1}_\text{\textsmaller{#2}}}
\newcommandx{\supm}[2][1=p,2=A,usedefault]{{#1}^{\!\mathsmaller{#2}}}
\newcommandx{\supt}[2][1=p,2=A,usedefault]{{#1}^\text{\textsmaller{#2}}}
\newcommandx{\subpt}[3][1=p,2=A,3=B,usedefault]{{#1}^\text{\textsmaller{#3}}_\text{\textsmaller{#2}}}
\newcommandx{\subpm}[3][1=p,2=A,3=B,usedefault]{{#1}^{\mathsmaller{#3}}_{\mathsmaller{#2}}}
\newcommandx{\LCTd}[4][1=\mu,2=\nu,3=\rho,4=\sigma,usedefault]{\veps_{#1#2#3#4}}
\newcommandx{\LCTu}[4][1=\mu,2=\nu,3=\rho,4=\sigma,usedefault]{\veps^{#1#2#3#4}}
\newcommandx{\gmetr}[2][1=\mu,2=\nu,usedefault]{g_{{#1}{#2}}}
\newcommandx{\invgmetr}[2][1=\mu,2=\nu,usedefault]{g^{{#1}{#2}}}
\newcommandx{\spc}[3][1=\mu,2=a,3=b,usedefault]{{\w_{#1}}^{\!\!{#2}{#3}}}
\newcommandx{\Conn}[3][1=\mu,2=\nu,3=\lambda,usedefault]{{\Gamma_{{#1}{#2}}}^{\!\!#3}}
\newcommandx{\viel}[2][1=\mu,2=a,usedefault]{{e_{#1}}^{\!#2}}
\newcommandx{\inviel}[2][1=a,2=\mu,usedefault]{{e_{#1}}^{#2}}
\newcommandx{\vieluu}[2][1=\mu,2=a,usedefault]{e^{#1#2}}
\newcommandx{\Rdduu}[4][1=\mu,2=\nu,3=a,4=b,usedefault]{{R_{{#1}{#2}}}^{{#3}{#4}}}

\newcommandx{\overbar}[1]{\mkern                   1.5mu\overline{\mkern-2.0mu#1\mkern-2.0mu}\mkern 1.5mu}
\newcommandx{\overbarcal}[1]{\mkern                   6.0mu\overline{\mkern-5.5mu#1\mkern-1.0mu}\mkern 1.5mu}  

\makeatletter
\normalsize
\makeatother

\DeclareFixedFont\trfont{OT1}{phv}{b}{sc}{11}

\title{%
       \centering\boldmath\LARGE\bfseries%
       New non-extremal and BPS hairy black holes in gauged $\,\mathcal{N}=2\,$ and $\,\mathcal{N}=8\,$ supergravity
       \vspace{1em}
       }

\usepackage[blocks]{authblk}  

\author[a,b]{Andres Anabalon%
    \thanks{{\href{mailto:andres.anabalon@uai.cl}{\texttt{andres.anabalon@uai.cl}}}}%
    }

\author[c]{Dumitru Astefanesei%
    \thanks{{\href{mailto:dumitru.astefanesei@pucv.cl}{\texttt{dumitru.astefanesei@pucv.cl}}}}%
    }

\author[d,e]{Antonio Gallerati%
    \thanks{{\href{mailto:antonio.gallerati@polito.it}{\texttt{antonio.gallerati@polito.it}}}}%
    }

\author[d,e]{Mario Trigiante%
    \thanks{{\href{mailto:mario.trigiante@polito.it}{\texttt{mario.trigiante@polito.it}}}}%
    }

\affil[a]{%
{Universidad Adolfo Ibáñez, Dep.\ de Ciencias, Fac.\ de Artes Liberales. Av.\ Padre Hurtado 750, Viña del Mar, Chile}\smallskip}

\affil[b]{%
{Departamento de Física, Universidad de Oviedo, Avda. Federico García Lorca 18, 33007 Oviedo, Spain}\smallskip}

\affil[c]{%
{Pontificia Universidad Católica de Valparaíso,
Instituto de Física, Av.\ Brasil 2950, Valparaíso, Chile}\smallskip}

\affil[d]{Politecnico di Torino, Dipartimento DISAT, corso Duca degli Abruzzi 24, 10129 Torino, Italy\smallskip}

\affil[e]{Istituto Nazionale di Fisica Nucleare (INFN), sez.\ TO, via Pietro Giuria 1, 10125 Torino, Italy}

\date{}

\makeatletter
\patchcmd{\@maketitle}{\begin{center}}{\begin{adjustwidth}{-0.25in}{-0.25in}\begin{center}}{}{}
\patchcmd{\@maketitle}{\end{center}}{\end{center}\end{adjustwidth}}{}{}
\makeatother

\begin{document}

\maketitle


\begin{abstract}
{\noindent\sloppy
In this article we study a family of four-dimensional, $\,\mathcal{N}=2$ supergravity theories that interpolates between all the single dilaton truncations of the $\SO(8)$ gauged $\mathcal{N}=8$ supergravity.
In this infinitely many theories characterized by two real numbers -- the interpolation parameter and the dyonic ``angle'' of the gauging -- we construct non-extremal electrically or magnetically charged black hole solutions
and their supersymmetric limits.\, All the supersymmetric black holes have non-singular horizons with spherical, hyperbolic or planar topology. Some of these supersymmetric and non-extremal black holes are new examples in the $\mathcal{N}=8$ theory that do not belong to the STU model. We compute the asymptotic charges, thermodynamics and boundary conditions of these black holes and show that all of them, except one, introduce a triple trace deformation in the dual theory.}
\end{abstract}

\bigskip

\tableofcontents


\bigskip


\section{Introduction} \label{sec:intro}
Anti-de Sitter (AdS) black hole solutions have received great attention in the last decades, due to the role they play in the phenomenology of the AdS/CFT conjecture \cite{Maldacena:1997re}. The latter is a holographic correspondence that relates  gravity theories in AdS spacetimes to dual field theories living in one dimension less. The classical AdS black hole solutions can provide relevant information about properties of the strongly coupled (dual) gauge theory and the duality can be also exploited to give a description of various condensed matter phenomena.\par
The study of thermodynamic properties of AdS black holes began with the seminal paper \cite{Hawking:1982dh}, where a first order phase transition from thermal AdS space to a (large) black hole phase was shown to exist. This analysis was subsequently extended to a variety of AdS black hole solutions \cite{Chamblin:1999tk,Chamblin:1999hg,Cvetic:1999ne,Caldarelli:1999xj}. These studies suggest that the black hole configurations possess phase structures that can reveal critical phenomena like many other common thermodynamic systems. In principle, the AdS/CFT dictionary should also allow to compute the microscopic entropy of AdS black holes and to compare it then with the macroscopic Bekenstein-Hawking result. This program has been rather successful in the supersymmetric case, see for instance \cite{Benini:2015eyy,Benini:2016rke}.
Indeed, of particular interest are black holes preserving a certain amount of supersymmetry allowing to map a weak (string) coupling computation of the entropy within superstring theory to the strong-coupling regime, where a formulation in terms of a black hole is valid \cite{Strominger:1996sh}.\par
The study of supersymmetric black hole solutions in  AdS started with the paper of Romans \cite{Romans:1991nq}, which showed that for minimal gauged $D=4$, \,$\mathcal{N}=2$ supergravity extremal, spherically symmetric black holes in four dimensions are not BPS. Romans also found a solution with hyperbolic horizon that is supersymmetric and he called it a ``cosmological dyon''. Later, it was shown that this solution represented a black hole, when the electric charge vanishes, with globally defined Killing spinors \cite{Caldarelli:1998hg}. This solution was then embedded in eleven dimensional supergravity in  \cite{Gauntlett:2001qs}.
We would like to remark that \cite{Cacciatori:2009iz} provided the first non-singular example of asymptotically AdS supersymmetric static black hole in four dimensions. This work then called for a comprehensive analysis of these solutions and their generalizations, see for instance \cite{Hristov:2010eu, Hristov:2010ri,Hristov:2011ye,Toldo:2012ec,Chow:2013gba,Gnecchi:2013mja, Gnecchi:2013mta, Lu:2014fpa, Faedo:2015jqa,Klemm:2015xda,Chimento:2015rra,Hristov:2018spe, Daniele:2019rpr}.\par\smallskip
The aim of this paper is the construction and discussion of new exact charged hairy black hole solutions in gauged $D=4$ supergravity, generalizing previous ones in \cite{Anabalon:2013eaa, Anabalon:2012ta, Anabalon:2017yhv,Anabalon:2020qux}. As we have discussed in the uncharged case in \cite{Anabalon:2017yhv, Anabalon:2019tcy}, the model of this paper interpolates between four single dilaton truncations of the $\SO(8)$, $\mathcal{N}=8$ supergravity \cite{deWit:1981sst,deWit:1982bul}, with a possible dyonic gauging \cite{DallAgata:2011aa,DallAgata:2012mfj,DallAgata:2012plb,DallAgata:2014tph,Inverso:2015viq}. These four truncations break $\SO(8)$ into $\SO(p)\times\SO(8-p)$. In the $\mathcal{N}=2$ case, one includes two $\U(1)$ vectors, and the truncations with even $p$ can be seen as a subsector of the STU model. The truncations with odd $p$ and two Maxwell fields were previously unknown. We therefore prove, by an explicit construction, that the case with $p=3$ and two vectors can be embedded in $\mathcal{N}=8$ supergravity. We leave the case with $p=1$ for a future work.\par
Generally, the infinite class of $\mathcal{N}=2$ theories are characterized by two continuous parameters, namely the interpolation parameter and a parameter associated to the dyonic gauging. We then provide explicit expressions for two new different families of non-extremal solutions and we analyse the duality relation between them. We also perform a systematic investigation of the thermodynamic properties of our new solutions and the analysis of the boundary conditions of the scalar field. It turns out that all these solution induce a triple trace deformation in the dual theory. Finally, we study the values of the parameters leading to supersymmetric configurations for our model. We also recover a known dyonic solution of the STU model which is a subcase of the general family found in \cite{Chow:2013gba}.\par
As an interesting historical remark, we note in appendix \ref{app:cases} that our non-extremal solutions, when constrained to the single dilaton truncations of the STU model, are the ones of Duff and Liu \cite{Duff:1999gh}. On the other hand, for the $T^3$ model our supersymmetric solutions are the ones of Cacciatori and Klemm \cite{Cacciatori:2009iz}. Hence, one can conclude that, for the purely electric and purely magnetic cases, the $T^3$ model Cacciatori-Klemm solutions are the non-singular BPS limit of Duff-Liu. \par
The paper is organized as follows. In Section \ref{sec:N2gaugedSUGRA} we review the general structure of $D=4$, \,$\mathcal{N}=2$ supergravity coupled to vector multiplets in the presence of FI terms. In Section \ref{sec:model} we focus on a specific class of models which interpolate between all the single dilaton truncations of the maximally supersymmetric $\SO(8)$  gauged, supergravity theory. Here we provide the black hole solutions in terms of canonically normalized fields, we study its thermodynamics and the boundary conditions. The two family of solutions are related by means of electromagnetic duality and we discuss this in detail. In Section \ref{sec:susybh} we consider, among these solutions, the BPS ones: we show that, independently of the topology of the horizon, the solutions have a non-singular BPS limit. The connection to known non-extremal and BPS solutions is presented in Appendix \ref{app:cases}. Finally, in Section \ref{sec:N8trunc}, we characterize certain models within the general class under consideration as consistent truncations of \,$\mathcal{N}=8$, $D=4$ gauged supergravities. Here we construct a previously unknown very simple truncation of the maximally supersymmetric theory. This allows to embed the corresponding black hole solutions in the four-dimensional $\mathcal{N}=8$ theory. Finally, we end with concluding remarks.

\section[\texorpdfstring{$\N=2$ gauged theory}{}]{\boldmath $\N=2$ gauged theory \unboldmath}
\label{sec:N2gaugedSUGRA}
Let us consider an extended \,$\N=2$ supergravity theory in four dimensions, coupled to $\nv$ vector multiplets and no hypermultiplets, in the presence of Fayet--Iliopoulos (FI) terms. The model describes $\nv+1$ vector fields $A^\Lambda_\mu$,\, ($\Lambda=0,\dots,\nv$) and $\ns=\nv$ complex scalar fields $z^i$ ($i=1,\dots,\ns$).
The bosonic gauged Lagrangian reads
\begin{equation}
\frac{1}{\sqrt{-g}}\,\Lagr_{\textsc{bos}}=
-\frac{R}{2}
\+g_{i\bar{\jmath}}\,\partial_\mu z^i\,\partial^\mu \zb^{\bar{\jmath}}
\+\frac{1}{4}\,\II_{\Lambda\Sigma}(z,\zb)\,F^\Lambda_{\mu\nu}\,F^{\Sigma\,\mu\nu}
\+\frac{1}{8\,\sqrt{-g} }\,\RR_{\Lambda\Sigma}(z,\zb)\,\veps^{\mu\nu\rho\sigma}\,F^\Lambda_{\mu\nu}\,F^{\Sigma}_{\rho\sigma}
\-V(z,\zb)\;,
\label{eq:boslagr}
\end{equation}
where $g=\det(g_{\mu\nu})$ and with the $\nv+1$ vector field strengths:
\begin{align}
F^\Lambda_{\mu\nu}\=\partial_\mu A^\Lambda_\nu-\partial_\nu A^\Lambda_\mu\;.
\end{align}
The $\ns$ complex scalars $z^i$ couple to the vector fields in a non-minimal way through the real symmetric matrices $\II_{\Lambda\Sigma}(z,\zb)$, $\RR_{\Lambda\Sigma}(z,\zb)$ and
span a special K\"ahler manifold $\Ms_\textsc{sk}$ (see App.\ \ref{app:geom} for definitions), while the scalar potential $V(z,\zb)$ originates from electric-magnetic FI terms.\par
In the following we will analyse a consistent dilaton truncation of a class of $\mathcal{N}=2$ supergravities. Within this model, we shall describe families of hairy black hole solutions, regular on and outside the horizon, with dyonic FI terms. These families are related by a symmetry transformation acting non-trivially also on the FI parameters of the model, thus providing a noteworthy solution generating technique in asymptotically AdS spacetimes%
\footnote{%
the assumption is not generic, since in asymptotically AdS black holes the solutions generating technique \cite{Breitenlohner:1987dg,Cvetic:1995kv,Cvetic:1996kv,Gaiotto:2007ag,Bergshoeff:2008be,Bossard:2009at,Fre:2011uy,Andrianopoli:2013kya,Andrianopoli:2013jra,Andrianopoli:2013ksa,Gallerati:2019mzs}, based on the global isometry group of the ungauged theory, can no longer be applied in a gauged model, due to the non-trivial duality action on the embedding tensor \cite{Trigiante:2016mnt,Gallerati:2016oyo}%
}.

\section{The model}\label{sec:model}
The framework under consideration is an $\mathcal{N}=2$ theory with no hypermultiplets and a single vector multiplet ($\nv=1$) containing a complex scalar field $z$.
The geometry of the special K\"ahler manifold (see App.\ \ref{subapp:geom}) is characterized by a prepotential of the form:
\begin{equation}
\mathcal{F}(\X^\Lambda)\:=-\frac{i}{4}\:\left(\X^0\right)^{n}\left(\X^1\right)^{2-n}\,,
\label{eq:prepotentialn}
\end{equation}
$\X^\Lambda(z)$ being components of a holomorphic section of the symplectic bundle over the manifold and the coordinate $z$ being identified with the ratio $\X^1/\X^0$, in a local patch in which $\X^0\neq 0$.\par
For special values of $n$, the model is a consistent truncation of the STU model. The latter is a $\N=2$ supergravity coupled to $\nv=3$ vector multiplets and characterized, in a suitable symplectic frame, by the prepotential:
\begin{equation}
\mathcal{F}_\textsc{stu}(\X^\Lambda)\:=-\frac{i}{4}\,\sqrt{\X^0\,\X^1\,\X^2\,\X^3}\;,
\end{equation}
and the scalar manifold is symmetric of the form $\mathscr{M}_\textsc{stu}=\big(\SL(2,\mathbb{R})/\SO(2)\big)^3$, spanned by the three complex scalars $z^i=\X^i/\X^0$, $i=1,2,3$. This model, for a certain choice of the FI terms, is in turn a consistent truncation of the maximal $\N=8$ theory in four-dimensions with ${\rm SO}(8)$ gauge group \cite{Duff:1999gh}.
For $n=1/2$, the theory under consideration is the so-called $z^3$--model, whose manifold is $\SL(2,\mathbb{R})/\SO(2)$ and is embedded in that of the STU model through the identification $z^1=z^2=z^3=z$.
Another consistent truncation of the the STU model is found for $n=1$: this theory is again characterized by a manifold of the form $\SL(2,\mathbb{R})/\SO(2)$ but with a different coupling of the scalar fields to the vectors and, in particular, it corresponds to the so-called \emph{minimal coupling model}, describing a $\N=2$ supergravity coupled to 1 abelian vector multiplet \cite{Luciani:1977hp}.
\par\smallskip
%
If we set $\X^0=1$, the holomorphic section $\Omega^M$ of the model reads:
\begin{equation}
\Omega^M=
\left(\begin{matrix}
1  \cr  z  \cr -\dfrac{i}{4}\,n\,z^{2-n} \cr -\dfrac{i}{4}\,(2-n)\,z^{1-n}
\end{matrix}\right)\;,
\end{equation}
and the K\"ahler potential $\mathcal{K}$ has the expression
\begin{equation}
e^{-\mathcal{K}}\=\frac{1}{4}\,z^{1-n}\,\big(n\,z-(n-2)\,\bar{z}\big)\+\text{c.c.}
\end{equation}
The theory is then deformed with the introduction of abelian electric-magnetic
FI terms defined by a constant symplectic vector $\theta^M=\left(\theta^1,\,\theta^2,\,\theta^3,\,\theta^4\right)$, encoding the gauge parameters of the model.
The scalar potential $V(z,\zb)$ can be then obtained from:
\begin{equation}
V\=\left(g^{i\bar{\jmath}}\,\Ucal_i^M\,\overbar{\Ucal}_{\bar{\jmath}}^N
         -3\,\V^M\,\overbar{\V}^N\right)\theta_M\,\theta_N\,=\,
    -\frac{1}{2}\,\theta_M\,\M^{MN}\,\theta_N-4\,\V^M\,\overbar{\V}^N\theta_M\,\theta_N\;,
\label{eq:Vpot}
\end{equation}
where \:$\V^M=e^{\mathcal{K}/2}\,\Omega^M$, \;$\Ucal_i^M=\D_i\,\V^M$\, and \,$\M(\phi)$\, is the symplectic, symmetric, negative definite matrix encoding the non-minimal couplings of the scalars to the vector fields of the theory (see App.\ \ref{subapp:geom}).\par\smallskip
Writing \,$z=e^{\lambda\,\phi}+i\,\chi$\,, the truncation to the dilaton field $\phi$
\begin{equation}
\chi=0\,,
\end{equation}
in the absence of electric and magnetic charges, is consistent provided:
\begin{equation}
(2-n)\,\theta_1\,\theta_3 - n\,\theta_2\, \theta_4\=0\:.
\label{eq:consisttrunc0}
\end{equation}
In the presence of electric and magnetic charges, sufficient conditions for a consistent truncation to the dilaton include, besides the above eq.\ \eqref{eq:consisttrunc0},
\begin{equation}
F^\Lambda_{\mu\nu}\:\partial_\chi\RR_{\Lambda\Sigma}\:{F}^{\Sigma}_{\rho\sigma}\:\varepsilon^{\mu\nu\rho\sigma}\big\rvert_{{}_{\chi=0}}=\:0\:.
\label{eq:consisttrunc1}
\end{equation}
Conditions \cref{eq:consisttrunc0,eq:consisttrunc1} come from the consistency of the axion field equation after the $\chi=0$ truncation (see also the scalars field equations described in Sect.\ \ref{subsec:Duality})%
\footnote{%
in the model under consideration, one has \,$\partial_\chi\II\rvert_{{}_{\chi=0}\!}\!=0$,\,
\,$\RR\rvert_{{}_{\chi=0}}\!=0$\,
and \,$\partial_\chi\RR\rvert_{{}_{\chi=0}}\!\neq0$}%
.\par\smallskip
The metric, restricted to the dilaton, reads:
\begin{equation}
ds^2\=2\,g_{z\bar{z}}\;dz\,d\bar{z}\,\big\vert_{{\!\!}_{\!\chi=0\atop d\chi=0}}
    =\:\frac{1}{2}\lambda^2\,n\,(2-n)\,d\phi^2\;,
\label{eq:modulimetric}
\end{equation}
and is positive provided $0<n<2$. Choosing
\begin{equation}
\lambda\=\sqrt{\frac{2}{n\,(2-n)}}\;,
\end{equation}
the kinetic term for $\phi$ is canonically normalized.
As a function of the dilaton only, the scalar potential has the following explicit form:
\begin{equation}
\begin{split}
V\left(\phi\right)\:=\,&-2\,e^{\lambda\,\phi\,(n-2)}\,\left(\frac{2\,n-1}{n}\,\theta_{1}^2
+4\,\theta_{1}\,\theta_{2}\;e^{\lambda\,\phi}
+\frac{2\,n-3}{n-2}\,\theta_{2}^2\;e^{2\,\lambda\,\phi}\right)-\\
&-\frac{1}{8}\;e^{-\lambda\,\phi\,(n-2)}\,\Big(\left(2\,n-1\right)\,n\,\theta_{3}^{2}
-4\,\theta_{3}\,\theta_{4}\,n\,\left(n-2\right)\,e^{-\lambda\,\phi}
+\left(n-2\right)\,\left(2\,n-3\right)\,\theta_{4}^{2}\;e^{-2\,\lambda\,\phi}\Big)\,.
\label{eq:potdil}
\end{split}
\end{equation}

\subsection{Redefinitions}\label{redefinitions}
Let us now make the shift
\begin{equation}
\phi\,\rightarrow\,\phi-\frac{2\,\nu}{\lambda\,(\nu+1)}\,\log(\theta_2\,\xi)\,,
\label{eq:phishift}
\end{equation}
and redefine the FI terms as:
\begin{equation} \label{eq:newpar}
\theta_{1}=\frac{\nu+1}{\nu-1}\;\theta_{2}^{-\frac{\nu-1}{\nu+1}}\,\xi^{-\frac{2\,\nu}{\nu+1}}\,,\qquad
\theta_{3}=2\,\alpha\left(\xi\,\theta_{2}\right)^{\frac{\nu-1}{\nu+1}}\,s\,,\qquad
\theta_{4}=\frac{2\,\alpha}{\theta_{2}\,\xi\,s}\,,
\end{equation}
having defined the quantity
\begin{equation}
\nu=\frac{1}{n-1}\:,
\end{equation}
and having also introduced the parameters $\alpha$, $\xi$ and $s$.
In light of the new parametrization \eqref{eq:newpar}, condition \eqref{eq:consisttrunc0} requires
\begin{equation}
\frac{2\,\alpha\,(s^2-1)\,(\nu+1)}{\nu\,\xi\,s}=0\;,
\end{equation}
which is solved, excluding the value $\nu=-1$ ($n=0$), either for pure electric FI terms ($\alpha=0$) or for $s=\pm1$. Since we are interested in dyonic FI terms, we shall restrict ourselves to the latter case.
For $s=\pm1$, it is straightforward to invert the the parameterization yielding:
\begin{equation}
\xi=\theta_2^{-1}\left(\frac{\theta_3}{\theta_4}\right)^{\!\tfrac{\nu+1}{2\,\nu}},\quad\qquad
\alpha^2=\frac{\theta_3\,\theta_4}{4}\left(\frac{\theta_3}{\theta_4}\right)^{\!\tfrac{1}{\nu}},
\end{equation}
which require that $\theta_3$ and $\theta_4$ have the same sign.\par\smallskip
Next, it is convenient to express $\xi$ in terms of the AdS radius $L$:
\begin{equation}
\xi\=\frac{2\,L\,\nu}{\nu-1}\,\frac{1}{\sqrt{1-\alpha^{2}\,L^{2}}}\;.
\end{equation}
This ensures the existence of an AdS vacuum without any further constraints on the original FI terms, as they just have to satisfy the inequality
\begin{equation}
\alpha^2\,L^2 \= \frac{(\nu-1)^2\:{\theta_3}^{\!2(\nu+1)/\nu}}%
    {(\nu-1)^2\:{\theta_3}^{\!2(\nu+1)/\nu}+\,16\,\nu^2\,{\theta_2}^{\!2}\,{\theta_4}^{\!2/\nu}}
    \;\leq\;1\;,
\end{equation}
which is always true. The only two points which are not possible to reach with our parameterization are either \,$\theta_3=0$ and $\theta_4\neq0$\, or \,$\theta_4=0$ and $\theta_3\neq0$.\par\smallskip
After the shift \eqref{eq:phishift}, the scalar field $z$ is expressed as
\begin{equation}
z\=\left(\theta_{2}\,\xi\right)^{-\frac{2\,\nu}{\nu+1}}\,e^{\lambda\,\phi}\,, \end{equation}
and the same redefinition for the potential (with $s=\pm1$) yields
\begingroup
\belowdisplayskip=0pt
\belowdisplayshortskip=0pt
\begin{equation}
\begin{split}
V(\phi)\:=\,&-\frac{\alpha^2}{\nu^2}\,\left(\frac{(\nu-1)(\nu-2)}{2}\,
    e^{-\phi\,\ell\,(\nu+1)} + 2\,(\nu^2-1)\,e^{-\phi\,\ell} +\frac{(\nu+1)(\nu+2)}{2}\,e^{\phi\,\ell\,(\nu-1)}\right)+
\\
    &+\frac{\alpha^2-L^{-2}}{\nu^2}\,\left(\frac{(\nu-1)(\nu-2)}{2}\,
    e^{\phi\,\ell\,(\nu+1)} + 2\,(\nu^2-1)\,e^{\phi\,\ell} +\frac{(\nu+1)(\nu+2)}{2}\,e^{-\phi\,\ell\,(\nu-1)}\right)\,,
\end{split}\label{VDmodel}
\end{equation}
\endgroup
where \,$\ell=\dfrac{\lambda}{\nu}$\, and having disposed of $\theta_2$ by the above redefinitions.\par\smallskip
After the restriction to the dilaton truncation, the matrix $\mathcal{R}_{\Lambda\Sigma}$ vanishes and the action has the form
\begin{equation}
\mathscr{S}\=\frac{1}{8\pi G}\int{
    d^4 x\;\sqrt{-g}  \left(-\,\frac{R}{2}\,+\,\frac{\partial_\mu\phi\,\partial^\mu\phi}{2}
    \,+\,\frac{1}{4}\:\mathcal{I}_{\Lambda\Sigma}(\phi)\,F^\Lambda_{\mu\nu}\,F^{\Sigma\,\mu\nu}
    -\,V(\phi)\right)} \;,\label{truncact}
\end{equation}
where the matrix $\mathcal{I}_{\Lambda\Sigma}$ explicitly reads
\begin{equation}
\mathcal{I}_{\Lambda\Sigma}\,=
    \left(\begin{array}{cc}
    -\tfrac{1+\nu}{4\,\nu}\,e^{(-1+\nu)\,\ell\,\phi}\,(\theta_2\,\xi)^{\frac{2\,(1-\nu)}{1+\nu}} & 0 \\
    0 & \tfrac{1-\nu}{4\,\nu}\,e^{-(1+\nu)\,\ell\,\phi}\,(\theta_2\,\xi)^2 \\
          \end{array}\right)\;.
\end{equation}
It is more transparent to work with of the canonically normalized gauge fields
\begin{equation}
\bar{F}^1=\,\frac{1}{2}\;\sqrt{\frac{1+\nu}{\nu}}\;(\theta_2\,\xi)^{\frac{1-\nu}{1+\nu}}\,F^1\:,
\qquad\quad
\bar{F}^2=\,\frac{1}{2}\;\sqrt{\frac{-1+\nu}{\nu}}\;(\theta_2\,\xi)\,F^2\:,\qquad
\label{cangauge}
\end{equation}
in terms of which the action reads
\begin{equation}
\mathscr{S}\=\frac{1}{8\pi G}\int{
    d^4 x\;\sqrt{-g}  \left(-\,\frac{R}{2}\,+\,\frac{\partial_\mu\phi\,\partial^\mu\phi}{2}
    \,-\,\frac{1}{4}\,e^{(-1+\nu)\,\ell\,\phi}\,\left(\bar{F}^1\right)^2 -\frac{1}{4}\,e^{-(1+\nu)\,\ell\,\phi}\,\left(\bar{F}^2\right)^2 -\,V(\phi)\right)} \;.
\label{eq:canonicact}
\end{equation}
%
%

\bigskip

\subsection{Hairy black hole solutions}
In what follows, we construct two distinct families of solutions, which we refer to as electric and magnetic, respectively. We present solutions with spherical, toroidal, and hyperbolic horizon topology and an analysis of their thermodynamics and asymptotic structure.

\subsubsection{Family 1 - Electric solutions}
A family of solutions is given by
\begingroup
\begin{subequations}\label{eq:elsol}
\begin{align}
&\phi\:=\,-\ell^{-1}\ln(x)\,,
\\[2ex]
&\bar{F}^1_{tx}\,=\:Q_1\,x^{-1+\nu},
\qquad\;\bar{F}^2_{tx}\,=\:Q_2\,x^{-1-\nu},
\\[2.75ex]
&f(x)\,=\:\frac{x^{2-\nu }\,\eta ^{2}\,(x^{\nu }-1)^{2}}{\nu ^{2}}\,\frac{%
k }{L^{2}}+\alpha ^{2}L^{2}\,\left( -1+\frac{x^{2}}{\nu ^{2}}\,\big(%
\,(\nu +2)\,x^{-\nu}-(\nu -2)\,x^{\nu }+\nu ^{2}-4\,\big)\right)+
\nonumber\\
    &\phantom{f(x)\,=\:} +1+\frac{x^{2-\nu}\,\eta^{2}\,(x^{\nu}-1)^{3}}{\nu^{3}L^{2}}\left(
    \frac{Q_{1}^{2}}{(\nu +1)}-\frac{Q_{2}^{2}}{(\nu-1)}\,x^{-\nu}\right) \;,
\\[1.75\jot]
&\Upsilon (x)\,=\:\frac{x^{\nu-1}\,\nu^{2}L^{2}}{\eta^{2}\,(x^{\nu}-1)^{2}}\,,
\\[1.75ex]
&ds^{2}\,=\;\Upsilon(x)\left(f(x)\,dt^{2}-\frac{\eta^{2}}{f(x)}\,dx^{2}
    -L^2\,d\Sigma_{k}\right)\,,\label{subeq:metric}
\end{align}
\end{subequations}
\endgroup
where $\eta$ is an integration constant, \:$d\Sigma_{k}^{2}=d\theta^2+\tfrac{\sin^2(\sqrt{k}\,\theta)}{k}\,d\varphi^2$\:
is the metric on the $2D$-surfaces $%
\Sigma _{k }=\{\mathbb{S}^{2},\,\mathbb{H}^{2},\,\mathbb{R}^{2}\}$
(sphere, hyperboloid and flat space) with constant scalar curvature
${R=2\,k}$.%
\footnote{%
To compare with our previous paper, \cite{Anabalon:2017yhv}, note that the scalar curvature of $\Sigma_{k}$ has a different normalization. This ammounts to $k^\textsc{here}=k^\textsc{there}\,L^2$.
}
The solutions are written in a form that makes the uncharged limit well-defined, yielding the hairy black holes of \cite{Anabalon:2017yhv}. This is not the standard use of the literature of static supergravity black holes, where the uncharged limit is either Schwarzschild or Schwarzschild-AdS spacetime.

\paragraph{Boundary conditions, mass and thermodynamics for the electric solutions.}
To compare with the AdS canonical coordinates, let us consider the following fall-off:
\begin{equation}
\Upsilon(x)\= \frac{r^2}{L^2} + O\left(r^{-2}\right)\;.
\end{equation}
The change of coordinates that provides the right asymptotic behaviour is
\begin{equation} \label{CC}
x\= 1 \pm \left(\frac{L^2}{\eta\,r}+L^6\,\frac{1-\nu^2}{24\,\left(\eta\,r\right)^3}\right)  +L^8\,\frac{\nu^2-1}{24\,\left(\eta\,r\right)^4}\;,
\end{equation}
where we take $\eta>0$ and the $\pm$ sign depends on whether one takes the $x<1$ ($-$) or $x>1$ ($+$). Accordingly, the asymptotic behaviour of the scalar field is
\begin{equation}\label{fall off}
\phi\=L^2\,\frac{\phi_0}{r}+L^4\,\frac{\phi_1}{r^2}+O\left(r^{-3}\right)\,=\,
    \mp L^2\,\frac{1}{\ell\,\eta\,r}+L^4\,\frac{1}{2\,\ell\,\eta^2\,r^2}+ O\left(r^{-3}\right)\;,
\end{equation}
where we have normalized $\phi_0$ and $\phi_1$ to match their conformal and engineering dimension. In the canonical coordinates, we can now easily read off the coefficients of the leading and subleading terms in the scalar boundary expansion
\begin{equation}
\label{f1}
\phi_0=\mp\frac{1}{\ell\,\eta}\;, \qquad
\phi_1=\frac{\ell}{2}\,\phi_{0}^2\;,
\end{equation}
which corresponds to AdS invariant boundary conditions, namely a triple trace deformation in the boundary theory. Hence, the boundary conformal symmetry is preserved and the black hole mass can be read-off from the asymptotic expansion of the spacetime \eqref{eq:elsol} \cite{Henneaux:2006hk, Anabalon:2014fla, Anabalon:2015xvl}. Indeed, let us consider the asymptotic expansion of the metric:
\begin{equation}
\begin{split}
g_{tt}&\=\frac{r^2}{L^2}+k-\frac{\mu_\textsc{e}\,L^4}{r}
    +O\left(r^{-2}\right)\;,\\[2\jot]
g_{rr}&\:=-\frac{L^2}{r^2}-L^6\:\frac{k\,L^{-2}+\frac{1}{2}\,\phi_0^2}{r^4}
          +O\left(r^{-5}\right)\;,
\end{split}
\end{equation}
where
\begin{equation}
\mu_\textsc{e}=\pm\left(\frac{\nu^2-4}{3\,\eta^3}\,\alpha^2\,L^2-\frac{k}{\eta\,L^2}
    +\frac{Q_2^2}{\eta\,(\nu-1)\,L^2}-\frac{Q_1^2}{\eta\,(\nu+1)\,L^2}\right)\;,
\end{equation}
and we have taken $\eta>0$. Thus, the black hole mass is
\begin{equation}
M_\tts{E}\=L^4\:\frac{\mu_\textsc{e}\,\sigma_k}{8\pi G}\;,
\end{equation}
where $\sigma_k=\int d\Sigma_k$. Concretely, the values are \,$\sigma_1=4\pi$\, and \,$\sigma_{-1}=8\pi (g-1)$,\, where \,$g \geq 2$\, is the genus of a compact negative constant curvature manifold. In the boundary, after discarding the conformal factor, the dual field theory lives
on a manifold of radius $L$ and therefore the energy density of a thermal gas of massless particles is \cite{Myers:1999psa}:
\begin{equation}
\label{rho}
\rho_\textsc{e}\=\frac{L^2}{8\pi G}\:\mu_\textsc{e}\;,
\end{equation}
while the temperature is given by
\begin{equation}
T\=\frac{\abs{f(x)'}}{4\pi\,\eta}\,\bigg|_{x=x_+}\:,
\end{equation}
where $f(x_+)=0$. Finally, the entropy is expressed as
\begin{equation}
S\=\frac{L^2\,\Upsilon(x_+)\,\sigma_k}{4\,G}\;.
\end{equation}
The physical charges and electric potentials are
\begin{equation}
\begin{split}
q_1&=\frac{L^2\,Q_1\,\sigma_k}{8\pi G\,\eta}\,,
\qquad
\Phi_1^\textsc{e}=Q_1\,\frac{x_+^\nu-1}{\nu}\,;
\\[2\jot]
q_2&=\frac{L^2\,Q_2\,\sigma_k}{8\pi G\,\eta}\,,
\qquad
\Phi_2^\textsc{e}=Q_2\,\frac{1-x_+^{-\nu}}{\nu}\:,
\end{split}
\end{equation}
and it is possible to verify that these quantities satisfy the first law of  thermodynamics
\begin{equation}
dM_\tts{E}\=T\,d S+\Phi_1^\textsc{e}\:d q_1+\Phi_2^\textsc{e}\:d q_2\;.
\end{equation}

\subsubsection{Family 2 - Magnetic Solutions}
A second family of solutions is given by
\begin{subequations} \label{eq:magsol}
\begin{align}
&\phi\:=\,+\,\ell^{-1}\ln (x)\,,
\\[1.5ex]
&\bar{F}^1_{\theta\varphi}\,=\:P_1\;\frac{\sin(\sqrt{k}\,\theta)}{\sqrt{k}}\;,
\qquad\;
\bar{F}^2_{\theta\varphi}\,=\:P_2\;\frac{\sin(\sqrt{k}\,\theta)}{\sqrt{k}}\;,
\\[2.5ex]
&f(x)\,=\:\frac{x^{2-\nu}\,\eta^2\,(x^\nu-1)^2}{\nu^2}\,
    \frac{k}{L^2}+\left(1-\alpha^2 L^2\right)\,\left(-1+\frac{x^{2}}{\nu^2}\,
    \big(\,(\nu +2)\,x^{-\nu }-(\nu -2)\,x^{\nu }+\nu^2-4\,\big)\right)
    +\nonumber \\
    &\phantom{f(x)\,=} +1+\frac{x^{2-\nu}\,\eta^4\,(x^\nu-1)^3}{\nu^3\,L^6}
    \left(\frac{P_{1}^2}{(\nu+1)}-\frac{P_{2}^2}{(\nu-1)}\,x^{-\nu}\right) \;,
\\[1.5\jot]
&\Upsilon(x)\,=\:\frac{x^{\nu-1}\,\nu^2\,L^2}{\eta^2\,(x^{\nu}-1)^2}\;,
\\[1.5ex]
&ds^{2}\,=\;\Upsilon(x)\left(f(x)\,dt^{2}-\frac{\eta^2}{f(x)}\,dx^2
    -L^2\,d\Sigma_k\right) \;.
\end{align}
\end{subequations}
The electric and magnetic solutions are related to each other by means of
electromagnetic duality
\begin{equation}
\phi\;\rightarrow\,-\phi\;,
\qqquad
\alpha^{2}\:\rightarrow\;L^{-2}-\alpha^{2}\;.
\end{equation}
and the corresponding transformation of the electromagnetic fields (see also Section\ \ref{subsec:Duality}). These two families of solutions are charged generalizations of the hairy black holes we have already obtained in \cite{Anabalon:2017yhv}. In each family, the asymptotic region is located at the pole of order 2 of the conformal factor $\Upsilon (x)$, namely $x=1$. The geometry and scalar field are singular at $x=0$ and $x=\infty$, therefore, the configuration contains two disjoint geometries given by $x\in(1,\infty)$ or $x\in(0,1)$.
The latter geometries are different, since the value (sign) of the scalar field $\phi$ is different depending whether $x$ is greater or less than 1, while the scalar potential is not an even function of $\phi$.%
\footnote{%
We discussed this in more detail in \cite{Anabalon:2017yhv}.
}\par
As we shall see in Sect.\ \ref{sec:susybh}, each of the two families of solutions is naturally defined in one of these disjoint regions. If the solution, described in one of these regions, is a regular black hole, then the region itself is a geodesically complete spacetime provided reflective boundary conditions are imposed at radial infinity.

\paragraph{Boundary conditions, mass and thermodynamics for the magnetic solutions.}
The metric of the magnetic family can be obtained from the electric solutions by means of the transformation
\begin{equation}
Q_i\,\rightarrow\,\frac{\eta}{L^2}\,P_i\;,
\qqquad\;
\alpha^{2}\:\rightarrow\;L^{-2}-\alpha^{2}\; .
\end{equation}
Therefore all the discussions on mass and thermodynamics carry on from the electric to the magnetic case. The only important difference is that the scalar field satisfies now other boundary conditions
\begin{equation}
\phi_1=-\frac{\ell}{2}\,\phi_{0}^2\;,
\end{equation}
which correspond again to AdS invariant boundary conditions, for more details see \cite{Anabalon:2017yhv}. We define now
\begin{equation}
\mu_\textsc{m}=\pm\left(\frac{\nu^2-4}{3\,\eta^3}\,(1-\alpha^{2}\,L^2)-\frac{k}{\eta\,L^2}
    +\frac{ \eta \, P_2^2}{(\nu-1)\,L^6}-\frac{\eta\, P_1^2}{(\nu+1)\,L^6}\right)\;,
\end{equation}
where the $\pm$ corresponds to the $x>1$ and $x<1$, respectively. The black hole mass is
\begin{equation}
M_\tts{M}\=L^4\:\frac{\mu_\textsc{m}\,\sigma_k}{8\pi G}\;,
\end{equation}
while the magnetic charges and potentials are
\begin{equation}
\begin{split}
p_1&=\frac{P_1\,\,\sigma_k}{8 \pi G}\,,
\qquad
\Phi_1^\textsc{m}=\frac{P_1\,\eta}{L^2}\,\frac{x_+^\nu-1}{\nu}\,;
\\[2\jot]
p_2&=\frac{P_2\,\sigma_k}{8 \pi G}\,,
\qquad
\Phi_2^\textsc{m}=\frac{P_2\,\eta}{L^2}\,\frac{1-x_+^{-\nu}}{\nu}\,.
\end{split}
\end{equation}
It is possible to verify that these quantities satisfy the first law of thermodynamics
\begin{equation}
dM_\tts{M}\=T\, dS+\Phi_1^\textsc{m}\:dp_1+\Phi_2^\textsc{m}\:dp_2\;.
\end{equation}

\subsection[\texorpdfstring{The special case \,$n=1$\: or \:$\nu=\infty$}{}]%
{\boldmath The special case \,$n=1$\: or \:$\nu=\infty$ \unboldmath}
This special class of solutions is interesting because it can be embedded in gauged $\N=8$ supergravity. The action in this case has the form
\begin{equation}
\mathscr{S}\=\frac{1}{8\pi G}\int{
    d^4 x\;\sqrt{-g}  \left(-\,\frac{R}{2}\,+\,\frac{\partial_\mu\phi\,\partial^\mu\phi}{2}
    \,-\,\frac{1}{4}\,e^{\sqrt{2}\,\phi}\left(\bar{F}^1\right)^2 \,-\,\frac{1}{4}\,e^{-\sqrt{2}\,\phi}\left(\bar{F}^2\right)^2
    \,+\,\frac{1}{L^2}\left(2+\cosh(\sqrt{2}\,\phi)\right)\right)}\:.
\end{equation}
The scalar potential coming from \eqref{eq:Vpot} in the $n=1$ limit is rewritten making the shift
\begin{equation}
\phi\,\rightarrow\,\phi-\sqrt{2}\,\log\left(\frac{\rho\,L}{2}\right)\,,
\end{equation}
and redefining the FI terms as:
\begin{equation}
\theta_{1}=\frac{\cos(\zeta)}{\rho\,L^2}\,,\qquad\;
\theta_{2}=\frac{\rho}{4}\,\cos(\zeta)\,,\qquad\;
\theta_{3}=\rho\,\sin(\zeta)\,,\qquad\;
\theta_{4}=\frac{4\,\sin(\zeta)}{\rho\,L^2}\,,
\end{equation}
while the field strengths have the form
\begin{equation}
\bar{F}^1=\,(\rho\,L)^{-1}\,F^1\,,
\qquad\quad
\bar{F}^2=\,\frac{\rho\,L}{4}\,F^2\,.\qquad
\end{equation}
The above Lagrangian yields a consistent truncation of the \,$\N=2$ minimal coupling supergravity (corresponding to $n=1$) provided the following constraint
\begin{equation}
\bar{F}^1\land\bar{F}^1\-e^{-2\,\sqrt{2}\,\phi}\,\bar{F}^2\land\bar{F}^2\=0
\label{cons}
\end{equation}
is satisfied. The above condition follows from eq.\ \eqref{eq:consisttrunc1} and is required for the consistent truncation to the dilaton only.\par
It is possible to obtain the following dyonic solution for this theory,
\begin{subequations}
\begin{align} \label{dyonic0}
&\phi\:=\,-\,\ln(x)/\sqrt{2}\,,
\\[1.5ex]
&\bar{F}^1_{tx}\,=\:Q_1\,,
    \qquad
    \bar{F}^1_{\theta\varphi}=\:P_1\,\frac{\sin(\sqrt{k}\,\theta)}{\sqrt{k}}\,,
\\[2\jot]
&\bar{F}^2_{tx}\,=\:\frac{Q_2}{x^2}\,,
    \qquad
    \bar{F}^2_{\theta\varphi}\,=\:P_2\,\frac{\sin(\sqrt{k}\,\theta)}{\sqrt{k}}\,,
\\[2.5ex]
&f(x)\,=\:\frac{k\,\eta^2\,(x-1)^2}{L^2\,x}+1+\frac{\eta^2\,(x-1)^3}{L^2\,x}\,
        \left(Q_1^2-\frac{Q_2^{2}}{x}-\frac{\eta^2}{L^4}\,\frac{P_1^2}{x}+\frac{\eta^2}{L^4}\,P_2^2\right)\,,\qquad\qquad
\\[\jot]
&\Upsilon (x)\,=\:\frac{x\,L^2}{\eta^2\,(x-1)^2}\;,
\\[1.5ex]
&ds^2\,=\:\Upsilon(x)\left(f(x)\,dt^{2}-\frac{\eta^2}{x^2\,f(x)}\,dx^{2}
    -L^2\,d\Sigma_k\right)\;,
\end{align}
\end{subequations}
and the constraint \eqref{cons} is solved if
\begin{equation}
P_1\,Q_1-P_2\,Q_2\=0\;.
\end{equation}
This dyonic solution is not new and it can be shown to be contained within the solutions of \cite{Chow:2013gba}. When an asymptotic analysis is done on these solutions, it can be found that the subleading term of the scalar field vanishes, $\phi_1=0$, and therefore only the leading term $\phi_0$ is non-trivial. This solution is also peculiar in that only the hyperbolic black hole has a non-singular supersymmetric limit. This can be easily verified with the expressions given in Section \ref{sec:susybh}.

\subsection{Duality relation between the two families of solutions}
\label{subsec:Duality}
The two families of solutions described earlier are related by a non-perturbative electric-magnetic duality symmetry. This duality is a global symmetry of the ungauged theory, namely of the model without the FI term $\theta_M$, and is extended to the gauged one once the constant tensor $\theta_M$ is made to transform under it as well. In general a transformation of $\theta_M$, being it a non-dynamical quantity, would imply a change in the theory and thus the duality would be interpreted as an equivalence between seemingly different models. Such transformation can be absorbed in a redefinition of a single parameter $\alpha$ in $\theta_M$: the two solutions related by duality satisfy the field equations of the same model with two dual values of the $\alpha$ parameter.\par
Let us first illustrate the action of the electric-magnetic duality equivalence among $\mathcal{N}=2$ supergravities with FI terms and then consider our specific model. Consider a generic $\mathcal{N}=2$ theory described by the Lagrangian:
\begin{equation}
\frac{1}{\sqrt{-g}}\,\Lagr_{\textsc{bos}}\:=
\-\frac{R}{2}
\+\frac{1}{2}\,\mathscr{G}_{rs}(\phi)\partial_\mu\phi^r\partial^\mu \phi^s
\+\frac{1}{4}\,\II_{\Lambda\Sigma}(\phi)\,F^\Lambda_{\mu\nu}\,F^{\Sigma\,\mu\nu}
\+\frac{1}{8\,\sqrt{-g} }\,\RR_{\Lambda\Sigma}(\phi)\,\veps^{\mu\nu\rho\sigma}\,F^\Lambda_{\mu\nu}\,F^{\Sigma}_{\rho\sigma}
\-V(\theta,\,\phi)\,,
\label{eq:boslagr0}
\end{equation}
where now we are describing the scalar manifold (which is complex of special K\"ahler type in the absence of hypermultiplets) in terms of real scalars that are collectively denoted by $\phi(x)\equiv (\phi^s(x))$, and where $\mathscr{G}_{rs}(\phi)$ is the metric on the scalar manifold in the chosen real parametrization. The scalar potential $V(\theta,\,\phi)$ is given by eq.\ \eqref{eq:Vpot}, and we have made explicit its dependence on both the scalar fields and the FI vector $\theta_M$. Here we shall focus only on the bosonic part of the action and of the field equations. To describe the duality it is useful to define the magnetic field strengths as follows:
\begin{equation}
G_{\Lambda\,\mu\nu}\,=
    -\,\epsilon_{\mu\nu\rho\sigma}\frac{\delta \Lagr_{\textsc{bos}}}{\delta{F}^\Lambda_{\rho\sigma}} \=\RR_{\Lambda\Sigma}\,{F}^\Sigma_{\mu\nu}-\II_{\Lambda\Sigma}\,\Hodge{\!F}^\Sigma_{\mu\nu}\,,
\end{equation}
where
\begin{equation}
\Hodge{\!F}^\Sigma_{\mu\nu}\,\equiv\, \frac{\sqrt{-g}}{2}\,\epsilon_{\mu\nu\rho\sigma}\,F^{\Lambda\,\rho\sigma}\,,
\end{equation}
and to introduce the symplectic field strength vector:
\begin{equation}
\FF^M=\,\left(\begin{matrix}F^\Lambda_{\mu\nu}\cr
    {G}_{\Lambda\,\mu\nu}\end{matrix}\right)\,.
\end{equation}
Defining the following $2(n_\text{v}+1)\times 2(n_\text{v}+1)$ matrices:
\begin{equation}
\mathcal{M}_{MN}(\phi)\,\equiv\,
    \left(\begin{matrix} \,\II_{\Lambda\Sigma}+(\RR\,\II^{-1}\,\RR)_{\Lambda\Sigma} & -(\RR\,\II^{-1})_{\Lambda}{}^\Gamma\cr -(\II^{-1}\,\RR)^\Delta{}_\Sigma & (\II^{-1})^{\Delta\Gamma}
    \end{matrix}\right),\qquad
\mathbb{C}\,\equiv\,
    \left(\begin{matrix} \Zero & \Id \cr -\Id & \Zero
    \end{matrix}\right),\qquad\label{MRI}
\end{equation}
where each block is $(n_\text{v}+1)\times (n_\text{v}+1)$,
the (bosonic part of) the field equations for the metric (Einstein equations), scalar fields and vector fields can be cast in the following compact form:
\begin{equation}
\begin{split}
\text{Einstein}:& \quad
     R_{\mu\nu}=\partial_\mu\phi^r\,\partial_\nu\phi^s\,\mathscr{G}_{rs}(\phi)+\frac{1}{2}\,\FF^M_{\mu\rho}\,\mathcal{M}_{MN}(\phi)\,\FF^N{\!}_\nu{}^\rho-V(\theta,\phi)\,g_{\mu\nu}\:,
\\
\text{scalars}:& \quad
    \mathcal{D}_\mu(\partial^\mu\phi^s)=\frac{1}{8}\,\mathscr{G}^{rs}(\phi)\,\FF^M_{\mu\nu}\:\partial_r\mathcal{M}_{MN}(\phi)\,\FF^{N\mu\nu}-\mathscr{G}^{rs}(\phi)\,\partial_rV(\theta,\phi)\:,
\\[\jot]
\text{vectors}:& \quad
    d\FF^M=0\:,\quad\;
    \Hodge\FF_{\mu\nu}=-\Cc\,\mathcal{M}(\phi)\,\FF_{\mu\nu}\:,
\label{eqsofmot}
\end{split}
\end{equation}
where, in the last equation (known as twisted self-duality condition), we have suppressed the symplectic indices $(M, N,\dots)$.\par
A feature of special K\"ahler manifolds, as well as of all scalar manifolds in extended supergravities with $\mathcal{N}>2$, is the existence of a flat symplectic bundle structure on it, within which the matrix $\mathcal{M}_{MN}$ can be regarded as a metric on the symplectic fiber. As a consequence of this, with each isometry ${\bf g}$ in the (identity sector of the) isometry group $G$ of the scalar manifold, there corresponds a constant $2(n_\text{v}+1)\times 2(n_\text{v}+1)$ symplectic matrix $\mathscr{R}[{\bf g}]=(\mathscr{R}[{\bf g}]^M{}_N)$ such that, if $\phi'=(\phi^{\prime s})=\phi'(\phi)$ describe the (non-linear) action of ${\bf g}$ on the scalar fields, we have that $\mathcal{M}_{MN}(\phi)$ transforms as a fiber metric:
\begin{equation}
\mathcal{M}(\phi')\=
    (\mathscr{R}[{\bf g}]^{-1})^T \mathcal{M}(\phi)\:\mathscr{R}[{\bf g}]^{-1}\:,
\end{equation}
where we have suppressed the symplectic indices. Being a symplectic matrix, $\mathscr{R}[{\bf g}]$ satisfies the condition $\mathscr{R}[{\bf g}]^T\,\mathbb{C}\,\mathscr{R}[{\bf g}]=\mathbb{C}$.
Then if, for any isometry ${\bf g}$, we transform the fields $\FF^M_{\mu\nu}$ and the constant vector $\theta_M$ correspondingly:
\begin{equation}
\FF^M_{\mu\nu}\,\rightarrow\:\FF^{\prime M}_{\mu\nu}=
    \mathscr{R}[{\bf g}]^M{}_N\,\FF^N_{\mu\nu}\:,
\qquad\quad
\theta_M\,\rightarrow\:\theta^{\prime}_M=
    (\mathscr{R}[{\bf g}]^{-1})^N{}_M\,\theta_N\:,
\end{equation}
the equations of motion (\ref{eqsofmot}) are formally left invariant. Indeed one can easily verify, for instance, that:
\begin{equation}
V(\theta',\phi')=V(\theta,\phi)\:,
\qquad\;
\FF^{\prime M}_{\mu\rho}\:\mathcal{M}_{MN}(\phi')\,\FF^{\prime N}{}_\nu{}^\rho=
    \FF^{M}_{\mu\rho}\:\mathcal{M}_{MN}(\phi)\,\FF^{N}{}_\nu{}^\rho\:.
\end{equation}
Aside from this duality equivalence, one can just redefine the field strengths and the FI constants without altering the physics of the model:
\begin{equation}
\FF_{\mu\nu}\,\rightarrow
    \left(\begin{matrix}{\bf A} & \Zero \cr \Zero & ({\bf A}^{-1})^T
    \end{matrix}\right)\FF_{\mu\nu}\:,
\qquad\quad
\theta\:\rightarrow\left(\begin{matrix} ({\bf A}^{-1})^T & \Zero \cr
        \Zero & {\bf A} \end{matrix}\right)\theta\:,
\label{redefs}
\end{equation}
where ${\bf A}=(A^\Lambda{}_\Sigma)$ is a generic real, invertible matrix, not related to the isometries of the scalar manifold. This latter corresponds to a linear redefinition of the vector fields and just amounts to choosing a different basis in the symplectic fiber.\par
The above mechanism also works for our 1-scalar truncated model, provided the isometry does not switch on the truncated axion (imaginary part of $z$). In this special case the indices $\Lambda,\,\Sigma$ run over two values, the scalar is only one and the metric is trivial ($\mathscr{G}(\phi)=1$). Moreover the absence of the axion also implies  $\RR_{\Lambda\Sigma}=0$, so that the matrix $\mathcal{M}_{MN}(\phi)$ has the simplified form:
\begin{equation}
\mathcal{M}_{MN}(\phi)\equiv
    \left(\begin{matrix} \,\II_{\Lambda\Sigma}(\phi)& \Zero \cr \Zero & (\II^{-1})^{\Delta\Gamma}(\phi)
\end{matrix}\right)\,.
\end{equation}
The isometries of the scalar field are
\begin{subequations}
\begin{align}
&\phi\,\rightarrow\:\phi'=\phi+\beta\,,\label{a}
\\[2\jot]
&\phi\,\rightarrow\:\phi'=-\phi\,,\label{b}
\end{align}
\end{subequations}
$\beta$ being a constant. The first of the above isometries were used in Section \ref{redefinitions} to reabsorb  the $\theta_2$ dependence of the FI terms and thus to make them simpler, basically only depending on the parameter $\alpha$. This was followed by a corresponding redefinition of the field strengths and of the FI parameters of the kind (\ref{redefs}). The resulting Lagrangian is (\ref{truncact}).\par
Let us now consider the duality transformation associated with the isometry (\ref{b}). Using the property that $\II_{\Lambda\Sigma}(-\phi)=(\II^{-1})^{\Lambda\Sigma}(\phi)$, the reader can easily verify that the duality transformation $\mathscr{R}[{\bf g}]$ associated with it is simply $\mathbb{C}$. Indeed we have:
\begin{equation}
\mathcal{M}(\phi')\=\mathcal{M}(-\phi)\=(\mathbb{C}^{-1})^T\mathcal{M}(\phi)\:\mathbb{C}^{-1}\:.
\end{equation}
By the same token we can compute the duality transformed field strengths:
\begin{equation}
\left(\begin{matrix}\bar{F}^{\prime \Lambda}_{\mu\nu}\cr \bar{G}'_{\Lambda\,\mu\nu}\end{matrix}\right)\,=\:
    \mathbb{C}\left(\begin{matrix}\bar{F}^{ \Lambda}_{\mu\nu}\cr
    \bar{G}_{\Lambda\,\mu\nu}\end{matrix}\right)\,=\,
    \left(\begin{matrix}\bar{G}_{\Lambda\,\mu\nu}\cr-\bar{F}^{ \Lambda}_{\mu\nu}\end{matrix}\right)\,.
\label{npemd}
\end{equation}
After the action of the shift symmetry (\ref{a}) and the above mentioned redefinition, the resulting FI parameter vector, to be denoted by  $\bar{\theta}_M$, has the form:
\begin{equation}
\bar{\theta}_M\,=\,\left(\sqrt{\frac{\nu+1}{\nu}} \sqrt{L^{-2}-\alpha^2}\,,\;\,
\sqrt{\frac{\nu-1}{\nu}} \sqrt{L^{-2}-\alpha^2}\,,\;\,
\alpha\,s\,\sqrt{\frac{\nu+1}{\nu}}\,,\;\,
\frac{\alpha}{s}\,\sqrt{\frac{\nu-1}{\nu}} \right)\,,
\end{equation}
We denote by $\bar{\theta}'$ the transformed of $\bar{\theta}$ by $\mathbb{C}$:
\begin{equation}
\bar{\theta}'\=\mathbb{C}\:\bar{\theta}\:,
\label{trathe}
\end{equation}
and one can verify that
\begin{equation}
V(\bar{\theta}',\phi')\=V(\bar{\theta},\phi)\:.
\end{equation}
Since $\bar{\theta}$ only depends on $\alpha$, we will simply denote the potential by $V(\alpha,\phi)$.
This action, apart from signs, basically amounts in $\bar{\theta}$ to changing $\alpha^2\rightarrow {L^{-2}-\alpha^2}$. In particular we find:
\begin{equation}
V(\alpha,\phi)\=V(\alpha',\phi')\:,
\label{Vinv}
\end{equation}
where $\alpha^{\prime\,2}={L^{-2}-\alpha^2}$.\par
Let us now prove the invariance of the equations of motion under the transformations (\ref{b}), (\ref{npemd}) and (\ref{trathe}). We start from the scalar field equation that reads:
\begin{equation}
\frac{1}{\sqrt{-g}}\,\partial_\mu(\sqrt{-g}\,\partial^\mu\phi)\=
    \frac{1}{4}\,\bar{F}_{\mu\nu}^\Lambda\,\partial_\phi \II_{\Lambda\Sigma}(\phi)\,\bar{F}^{\Sigma\,\mu\nu}-\partial_\phi V(\alpha,\,\phi)\,.\label{eqphi}
\end{equation}
We can dualize the vector fields into their magnetic duals, whose field strengths are $\bar{G}_{\Lambda\,\mu\nu}$. The field equations in terms of the dual field strengths are obtained from (\ref{eqphi}) simply by expressing, in the equation, $\bar{F}_{\mu\nu}^\Lambda$ in terms of $\bar{G}_{\Lambda\,\mu\nu}$, that is
$\bar{F}_{\mu\nu}^\Lambda=(\II^{-1})^{\Lambda\Sigma}(\phi)\,\Hodge\bar{G}_{\Sigma\,\mu\nu}$.
We then find:
\begin{equation}
\frac{1}{\sqrt{-g}}\,\partial_\mu(\sqrt{-g} \,\partial^\mu\phi)\=
    \frac{1}{4}\,\bar{G}_{\Lambda\mu\nu}\,\partial_\phi(\II^{-1})^{\Lambda\Sigma}(\phi)\:\bar{G}_\Sigma^{\mu\nu} -\partial_\phi V(\alpha,\,\phi)\:.
\label{eqphim}
\end{equation}
This equation can be derived from a ``dual action'' obtained by adding to (\ref{truncact}) the term $-\tfrac{1}{4}\,\bar{F}^\Lambda_{\mu\nu}\,\bar{G}_{\Lambda\,\rho\sigma}\,\epsilon^{\mu\nu\rho\sigma}$ and intergating out $\bar{F}^\Lambda_{\mu\nu}$ in favor of their duals. We can now easily verify that, if the electric solutions
\begin{equation}
\phi(x)\,,\;\;\bar{F}_{\mu\nu}^\Lambda(x)\,,\;\;g_{\mu\nu}(x)\,,
\label{originalelectric}
\end{equation}
solve the field equations with FI terms defined by the parameter $\alpha$, then the magnetic configuration
\begin{equation}
\phi'(x)=-\phi(x)\,,\quad\bar{G}_{\Lambda\,\mu\nu}^{\prime}(x)=-\bar{F}^\Lambda_{\mu\nu}(x)\,,\quad g'_{\mu\nu}(x)=g_{\mu\nu}(x)\,,
\label{newmagnetic}
\end{equation}
is a solution with parameter $\alpha'$. Indeed let us write (\ref{eqphim}) in terms of the primed quantities:
\begin{equation}
\frac{1}{\sqrt{-g}}\,\partial_\mu(\sqrt{-g}\,\partial^\mu\phi')\=
    \frac{1}{4}\,\bar{G}'_{\Lambda\mu\nu}\,\partial_{\phi'}(\II^{-1})^{\Lambda\Sigma}(\phi')\:\bar{G}_\Sigma^{\prime\mu\nu} -\partial_{\phi'} V(\alpha',\phi')\:.
\label{eqphiprime}
\end{equation}
Replacing the primed quantities in terms of the unprimed ones, the left hand side together with the derivatives with respect to the scalar on the right hand side, change sign. Using (\ref{Vinv}) and the property $\II_{\Lambda\Sigma}(-\phi)=(\II^{-1})^{\Lambda\Sigma}(\phi)$ the above equation becomes (\ref{eqphi}) in the fields (\ref{originalelectric}), which is satisfied by assumption. The identification $\bar{G}_{\Lambda\,\mu\nu}^{\prime}(x)=-\bar{F}^\Lambda_{\mu\nu}(x)$ means that the magnetic field strengths (apart from a sign) will have the same form as the original electric ones, so that we can relate its magnetic charge parameters $P_i'$ with the original electric ones: $P_i'=-Q_i$. The magnetic solution, has been described in terms electric field strengths $\bar{F}^{\prime\Lambda}_{\mu\nu}$, which are given by:
\begin{equation}
\bar{F}^{\prime\Lambda}_{\mu\nu}\=(\II^{-1})^{\Lambda\Sigma}(\phi')\:\frac{\sqrt{-g} }{2}\,\epsilon_{\mu\nu\rho\sigma}\,\bar{G}_{\Sigma}^{\prime\,\rho\sigma}=
    -\II_{\Lambda\Sigma}(\phi)\:\frac{\sqrt{-g} }{2}\,\epsilon_{\mu\nu\rho\sigma}\,\bar{F}^{\Sigma\,\rho\sigma}\:.
\end{equation}
By the same token one can also prove that the Einstein equations in the transformed fields of the magnetic background are just the same equations in the original ones, which is satisfied.
The metric is a duality invariant field, see the last of eqs. (\ref{newmagnetic}). Therefore $g_{\mu\nu}$ in the new solution is simply obtained from that in the original electric one by expressing $Q_i$ in terms of $P'_i$ and $\alpha$ in terms of $\alpha'$.

\subsection{Supersymmetric solutions}
Now we want to study supersymmetryc configurations for our model, namely configurations satisfying eqs.\ \eqref{eq:susyvar} \cite{Gallerati:2019mzs}. First it is useful to make a change of coordinates that puts metric \eqref{subeq:metric} in the standard form (see App.\ \ref{subapp:act})
\begin{equation}
ds^2\=e^{2\,U(r)}\,dt^2-e^{-2\,U(r)}\left(dr^2+e^{2\,\Psi(r)}\,d\Sigma_k^2\right)\;.
\end{equation}
This can be achieved through the change of coordinate
\begin{equation}
x(r)\=\left(1+\frac{L^2\,\nu}{\eta\,(r-c)}\right)^\frac{1}{\nu}\,,
\end{equation}
$c$ being a constant.%
\footnote{%
The new metric functions $e^{2U}$, $e^{2\Psi}$ are readily found to be \,${e^{2\,U(r)}=\Upsilon(x)\,f(x)\big\rvert_{{x=x(r)}}}$\, and
\,${e^{2\,\Psi(r)}=\Upsilon^2(x)\,f(x)\,L^2\big\rvert_{{x=x(r)}}}$.
}

\subsubsection{Family 1}
The scalar field $z$ in the new parametrization has the form
\begin{equation}
z\=(\theta_2\,\xi)^{-\frac{2\,\nu}{1+\nu}}\left(1+\frac{L^2\,\nu}{\eta\,(r-c)}\right)^{-1}\,,
\end{equation}
while the electric-magnetic charges explicitly read
\begin{equation}
\Gamma^M\=\left(\begin{array}{c}
                m^\Lambda \\
                e_\Lambda \\
                \end{array}\right)
    \=\left(
        \begin{array}{c}
          0 \\
          0 \\[1.5ex]
          \dfrac{L^2}{2\,\eta}\,Q_1\,\sqrt{\frac{1+\nu}{\nu}}\:(\theta_2\,\xi)^{\frac{1-\nu}{1+\nu}}\\[2.5ex]
          \dfrac{L^2}{2\,\eta}\,Q_2\,\sqrt{\frac{-1+\nu}{\nu}}\;\theta_2\,\xi\\
        \end{array}
      \right)\;.
\end{equation}
The solution is supersymmetric if
\begin{equation}\label{eq:BPSel}
\begin{split}
Q_1&=-\,Q_2\;\sqrt{\frac{-1+\nu}{1+\nu}}\,+\,\frac{k\,\eta}{\alpha\,L^2}\:\sqrt{\frac{\nu}{1+\nu}}\:,
\\[1.5ex]
Q_2&=\left(\frac{k\,\eta}{2\,\alpha\,L^2}+\frac{\alpha\,L^2\,(1+\nu)}{2\,\eta}\right)\,\sqrt{\frac{-1+\nu}{\nu}}\;.
\end{split}
\end{equation}

\subsubsection{Family 2}
The scalar field $z$ reads
\begin{equation}
z\=(\theta_2\,\xi)^{-\frac{2\,\nu}{1+\nu}}\left(1+\frac{L^2\,\nu}{\eta\,(r-c)}\right)\,,
\end{equation}
while the electric-magnetic charges have the form
\begin{equation}
\Gamma^M\=\left(\begin{array}{c}
                m^\Lambda \\
                e_\Lambda \\
                \end{array}\right)
    \=\left(
        \begin{array}{c}
        2\,P_1\,\sqrt{\frac{\nu}{1+\nu}}\:(\theta_2\,\xi)^{\frac{-1+\nu}{1+\nu}}\\[1.5ex]
        2\,P_2\,\sqrt{\frac{\nu}{-1+\nu}}\:(\theta_2\,\xi)^{-1}\\[1.5ex]
        0 \\
        0 \\
        \end{array}
      \right)\;.
\end{equation}
The solution is supersymmetric if
\begin{equation}
\begin{split}
P_1&=-\,P_2\;\sqrt{\frac{-1+\nu}{1+\nu}}\,+\,\frac{k\,L}{\sqrt{1-\alpha^2\,L^2}}\;\sqrt{\frac{\nu}{1+\nu}}\:,
\\[1.5ex]
P_2&=\left(\frac{k\,L}{2\,\sqrt{1-\alpha^2\,L^2}}+\frac{L^3\,(1+\nu)}{2\,\eta^2}\:\sqrt{1-\alpha^2\,L^2}\right)\,\sqrt{\frac{-1+\nu}{\nu}}\;.
\end{split}
\label{BPS magnetic}
\end{equation}
The supersymmetric magnetic condition can be obtained from the supersymmetric electric condition by means of the duality transformation
\begin{equation}
Q_i\rightarrow\,\frac{\eta}{L^2}\,P_i\;,
\qqquad
\alpha^{2}\:\rightarrow\;L^{-2}-\alpha^{2}\:.
\end{equation}

\subsubsection[\texorpdfstring{Special case \,$n=1$}{}]%
{\boldmath Special case \,$n=1$ \unboldmath}
The change of coordinate in this special case is given by
\begin{equation}
x(r)\=1+\frac{L^2}{\eta\,(r-c)}\:.
\end{equation}
The scalar field $z$ in the new parametrization has the form
\begin{equation}
z\=\frac{4}{(\rho\,L)^2}\left(1-\frac{L^2}{L^2+\eta\,(r-c)}\right)\,,
\end{equation}
while the electric-magnetic charges explicitly read
\begin{equation}
\Gamma^M\=\left(\begin{array}{c}
                m^\Lambda \\
                e_\Lambda \\
                \end{array}\right)
    \=\left(
        \begin{array}{c}
          P_1\,\rho\,L \\[1ex]
          P_2\,\dfrac{4}{\rho\,L} \\[2.5ex]
          Q_1\,\dfrac{L}{\rho\,\eta}\\[2.5ex]
          Q_2\,\dfrac{\rho\,L^3}{4\,\eta}\\
        \end{array}
      \right)\;.
\end{equation}
The solution is supersymmetric if
\begin{equation}\label{eq:BPSn1}
Q_1=\,Q_2\,,\qquad\;\;
P_1=P_2\,,\qquad\;\;
P_2=\frac{k\,L}{2\,\cos(\zeta)}-Q_2\,\frac{L^2\,\tan(\zeta)}{\eta}\,.
\end{equation}

\section{Supersymmetric black holes of finite area}\label{sec:susybh}
\subsection{Family 1: BPS electric black holes}
We have found that this family has BPS black holes of finite area only when $\alpha^{2}=L^{-2}$, namely when the gauging is purely magnetic. In this case the lapse function has a double zero as expected
\begin{equation}
\begin{split}
&f(x)=\:\frac{x^{2-2\,\nu}}{4\,L^{4}\,\nu^4}\,
    \Big((x^\nu-1)^2\,k\,\eta^2+L^{2}\big(2\,x^\nu(\nu^{2}-1)+x^{2\,\nu}(1-\nu)+\nu +1\big)\Big)^2\,,
\\[2ex]
&f\left(x_{+}\right)=0 \quad\Longrightarrow\quad\,
    x_{\pm}^{\nu}=\frac{\nu^2-1-k\,\eta^2\,L^{-2}}{\nu-1-k\,\eta^2\,L^{-2}}
    \pm\nu\,\frac{\sqrt{\nu^2-1-2\,k\,\eta^2\,L^{-2}}}{\nu-1-k\,\eta^2\,L^{-2}}\;.
\end{split}
\end{equation}
It is possible to verify that $x_\pm(\nu)=x_\pm(-\nu)$, namely, $x_\pm$ is an even function of $\nu$. Hence, without loss of generality is possible to make the analysis on the existence of horizons under the assumption that $\nu\geq 1$. Let us remark that the asymptotic region of the spacetime is located at $x=1$. Therefore, the region of the spacetime at $x>1$ is disconnected from the
region at $x<1$ and they represent different spacetimes.

\paragraph{Planar black holes.} When the horizon is flat ($k=0$) the location of the horizon is very simple
\begin{equation}
x_{\pm}^\nu\=\nu+1 \,\pm\, \nu\,\frac{\sqrt{\nu^2-1}}{\nu -1}\;,
\end{equation}
and it follows that $x^\nu_{+}>0$ and $x^\nu_{-}<0$ . We conclude that only $x_+$ exists.

\paragraph{Locally hyperbolic black holes.}
When $k =-1$, $x_{+}>1$ always exists while $0<x_{-}<1$ exists provided
\begin{equation}
x_{-}>0 \quad\Leftrightarrow\quad \eta^{2}\,L^{-2}>\nu+1\;.
\end{equation}

\paragraph{Spherically symmetric black holes.}
For spherical black holes, $k =1$, only $x_{+}$ exists:
\begin{equation}
\infty\,>\,x_{+}\,>\,\nu+1+\nu\,\sqrt{\frac{\nu+1}{\nu-1}}
\quad\Leftrightarrow\quad
\nu-1\,>\,\eta^{2}\,L^{-2}\,>\,0\;.
\end{equation}

\paragraph{Supersymmetric thermodynamics.}
From the BPS conditions \eqref{eq:BPSel} we find that the physical charges and electric potentials for the supersymmetric black hole solutions are
\begin{equation}
\begin{split}
q_1^\textsc{bps}&=\pm \frac{L^2\sigma_k}{8\pi G}
    \left(\frac{k}{2\,L}+\frac{L\,(1-\nu)}{2\,\eta^2}\right)\,\sqrt{\frac{\nu+1}{\nu}}\:,
\\[\jot]
\Phi_1^\textsc{bps}&=\pm\left(\frac{k\,\eta}{2\,L}+\frac{L\,(1-\nu)}{2\,\eta}\right)
    \,\frac{x_+^\nu-1}{\nu}\:\sqrt{\frac{\nu+1}{\nu}}\:,
\\[2\jot]
q_2^\textsc{bps}&=\pm\frac{L^2\sigma_k}{8\pi G}
    \left(\frac{k}{2\,L}+\frac{L\,(1+\nu)}{2\,\eta^2}\right)\,\sqrt{\frac{\nu-1}{\nu}}\:,
\\[\jot]
\Phi_2^\textsc{bps}&=\pm \left(\frac{k\,\eta}{2\,L}+
    \frac{L\,(1+\nu)}{2\,\eta}\right)\,\frac{1-x_+^{-\nu}}{\nu}\,\sqrt{\frac{\nu-1}{\nu}}\:,
\end{split}
\end{equation}
where we have set $\alpha=\pm\,L^{-1}$. The mass of the electric BPS black holes is given by
\begin{equation}
M_\tts{E}^\textsc{bps}\=\frac{L^4\,\sigma_k}{8 \pi G}
    \,\frac{\nu^2-1}{3\,\eta^3}\:.
\end{equation}
As expected, these quantities indeed satisfy the extremality condition
\begin{equation}
\delta M_\tts{E}^\textsc{bps}\=\Phi_1^\textsc{bps}\,\delta q_1^\textsc{bps}
    \,+\:\Phi_2^\textsc{bps}\,\delta q_2^\textsc{bps}\:.
\end{equation}

\subsection{Family 2: BPS magnetic black holes}
Again, we shall consider only the case  $\nu >1$. This family has BPS\
black holes of finite area only when $\alpha^2=0,$ namely when the
gauging is purely electric. In this case the metric of the magnetic solution exactly coincides with the metric of the electric solutions. Hence the analysis of the location of the horizons is exactly the same. The extremality of the magnetic solutions is the same as for the electric solutions when the electric charges and potentials are interchanged by their magnetic counterparts.

\section{$\mathcal{N}=8$ truncations}  \label{sec:N8trunc}
In this Section we briefly discuss the embedding of our models within maximal four-dimensional supergravity.%
\footnote{Only in this section we shall denote by $\Lambda,\Sigma$ the indices labelling the 28 vectors of the maximal theory; by $\lambda,\sigma=1,2$ those of the two  vectors surviving the truncation; by $M,N$ the symplectic indices of the 56 electric and magnetic charges; by $m,n=1,\dots, 4$ the symplectic index labelling the two electric and two magnetic charges in the truncated model. The corresponding indices in the new, $\omega$-rotated symplectic frame are distinguished from those in the original frame by a hat.}

\subsection{Uncharged case}
The original ${\rm SO}(8)$ gauging of $D=4$, \,$\mathcal{N}=8$ supergravity \cite{deWit:1981sst,deWit:1982bul} and its generalizations to non-compact/non-semisimple gauge groups ${\rm CSO}(p,q,r)$, $p+q+r=8$, \cite{Hull:1984yy,Hull:1984vg}, were recently shown to be part of a much broader class of gauged maximal theories
\cite{DallAgata:2011aa,DallAgata:2012mfj,DallAgata:2012plb,DallAgata:2014tph,Inverso:2015viq}, also, somewhat improperly, referred to as ``dyonic'' gaugings, see \cite{Trigiante:2016mnt,Gallerati:2016oyo} for a review.
The infinitely many theories we have described so far contain all the possible one-dilaton consistent truncations of the $\omega$-deformed ${\rm SO}(8)$ gauged maximal supergravities. To see this let us recall the main facts about these theories which are relevant to our discussion.\par
The field content of maximal supergravity consists of the graviton, 8 gravitini, 28 vector fields, 56 spin $1/2$ fields and 70 scalars. In the dyonic models with gauge group ${\rm SO}(8)$, the 28 generators of this group are gauged by the 28 vector fields in a symplectic frame which is related to the one of \cite{deWit:1981sst,deWit:1982bul}, in which the group ${\rm SL}(8,\mathbb{R})$ has a diagonal symplectic embedding, by an ${\rm SO}(2)$ transformation parametrizad by an angle $\omega$. The physically independent values of $\omega$ lie within the interval $0\le \omega\le \pi/8$, \,$\omega=0$ corresponding to the original theory by de Wit and Nicolai.\par
The 70 scalar fields parameterize the coset $\mathrm{E}_{7(7)}/{\rm SU}(8)$. One can choose a parametrization of this manifold which is covariant under ${\rm SU}(8)$. With this choice the scalars split into the representations ${\bf 35}_\text{c}$ and ${\bf 35}_\text{v}$ of the gauge group ${\rm SO}(8)$. The former can be truncated out, while the latter scalars span the submanifold $\mathrm{SL}(8,\mathbb{R})/\mathrm{SO}(8)$. The local $\mathrm{SO}(8)$ transformations can be used to diagonalize the coset representative and thus to further truncate the theory to the seven scalars parameterizing the non compact Cartan subalgebra of $\mathrm{SL}(8,\mathbb{R})$.%
\footnote{\label{so8}%
To see this it is useful to adopt a parametrization of the coset $\mathrm{SL}(8,\mathbb{R})/\mathrm{SO}(8)$ in terms of 28 Euler angles $(\xi_\alpha)$ parametrizing ${\rm SO}(8)$ and seven independent non-compact ``radii'' $\vec{\phi}=(\phi_i)$, with $i=1,\dots, 8$ and $\sum_{i=1}^8\phi_i=0$. This amounts to choosing a coset prepresentative for the manifold of the form $\mathbb{L}=\mathbb{L}_{{\rm SO}(8)}\,\mathbb{L}_{\mathcal{C}}$, where $\mathbb{L}_{{\rm SO}(8)}(\xi)$ is an element of ${\rm SO}(8)$ while $\mathbb{L}_{\mathcal{C}}(\vec{\phi})$ belongs to $e^{\mathcal{C}}={\rm O}(1,1)^7$. The factor  $\mathbb{L}_{{\rm SO}(8)}(\xi)$ can be disposed of by a gauge transformation: $\mathbb{L}\rightarrow \mathbb{L}'=\mathbb{L}_{\mathcal{C}}$, \:$A_\mu^\Lambda T_\Lambda\rightarrow\,A_\mu^{\prime \Lambda} T_\Lambda=A_\mu^\Lambda\mathbb{L}_{{\rm SO}(8)}^{-1}T_\Lambda \mathbb{L}_{{\rm SO}(8)}-\frac{1}{g}\,\mathbb{L}_{{\rm SO}(8)}^{-1}\partial_\mu \mathbb{L}_{{\rm SO}(8)}$, where $T_\Lambda$ are the ${\rm SO}(8)$ gauge generators, $\Lambda=1,\dots, 28$. By further setting $A_\mu^{\prime \Lambda} =0$ we truncate the bosonic sector to the seven scalar fields $\phi_i$ only.}
Upon truncation to gravity and scalar field sector, we are led to consider the following action \cite{Cvetic:1999xx, Cvetic:2000eb}:
\begin{equation}
I\big(g_{\mu\nu},\vec{\phi}\,\big) \=
    \int_{\mathcal{M}} \!\!d^{4}x\:\sqrt{-g}\:
    \left[-\frac{R}{2}+\frac{1}{2}\,\big(\partial\vec{\phi}\,\big)^2
    -V\big(\vec{\phi}\,\big)\right]\:,
\label{actionCv}
\end{equation}
where $\vec{\phi}=(\phi_i)$, with $i=1,\dots,8$\, and \,$\sum_{i=1}^8\phi_i=0$. The potential is given by
\begin{equation}
V\big(\vec{\phi}\,\big)\:=\,-\,\frac{g^2}{32}\,\left[\cos^2(\omega)\left(
\Big(\sum_{i=1}^8 X_i\Big)^2-2\,\sum_{i=1}^8 X_i^2\,\right)+\sin^2(\omega)\left(
\Big(\sum_{i=1}^8 X_i^{-1}\Big)^{2}-2\,\sum_{i=1}^8 X_i^{-2}\,\right)\right]\:,
\end{equation}
where
\begin{equation}
X_i=e^{2\,\phi_i}\:,
\qquad\quad
\prod_{i=1}^8\,X_i = 1\:.
\end{equation}
Let us consider now a single scalar field reduction preserving $\mathrm{SO}(p)\times\mathrm{SO}(8-p)$. This is effected through the following identification:
\begin{equation}
\phi_1=\dots=\phi_p=\frac{1}{2\sqrt{2}}\,\sigma\,\phi\,,
\qqquad\;
\phi_{p+1}=\dots=\phi_8=-\frac{1}{2\sqrt{2}}\,\frac{\phi}{\sigma}\,,
\end{equation}
where we have defined:
\begin{equation}
\sigma =\sqrt{\frac{8-p}{p}}= \sqrt{\frac{\nu-1}{\nu+1}}\,, \qqquad
p = \frac{4\,(\nu+1)}{\nu}\:.
\end{equation}
With the above choice we have:
\begin{equation}
X_1 = \cdots = X_p = X := e^{\frac{1}{\sqrt{2}}\,\sigma\,\phi}\:, \qquad\;\;
X_{p+1} = \cdots = X_8 = Y
    := e^{- \frac{1}{\sqrt{2}}\,\frac{\phi}{\sigma}}\:,\label{Xphi}
\end{equation}
and the previous action \eqref{actionCv} reduces to the one we are studying in this paper. The action is invariant under $\sigma \to 1/\sigma$, \,$\phi \to -\phi$\, and \,$p\to 8-p$. This action, consistent truncation of the $\omega$-rotated ${\rm SO}(8)$-gauged maximal supergravity, coincides, in the absence of vector fields, with the action (\ref{truncact}) upon the following identification:
\begin{equation}
g=\frac{\sqrt{2}}{L}\:,\qquad\;
\cos(\omega)=L\,\alpha\:,\qquad\;
\sin(\omega)=\sqrt{1-L^2\,\alpha^2}\:.
\end{equation}
Changing $\phi$ into $-\phi$ in (\ref{Xphi}), the above identifications change correspondingly:
\begin{equation}
g=\frac{\sqrt{2}}{L}\:,\qquad\;
\sin(\omega)=L\,\alpha\:,\qquad\;
\cos(\omega)=\sqrt{1-L^2\,\alpha^2}\:.
\end{equation}
From the above relation between $p$ and $\nu$, we conclude that the single scalar field models considered in this work, if all vector fields are set to zero, coincide with truncations of the $\omega$-deformed ${\rm SO}(8)$ gauged maximal supergravity to the singlet sector with respect to the following subgroups of  ${\rm SO}(8)$ gauge group :
\begin{equation}
\begin{tabular}[m]{rcl}%
$\nu=\frac{4}{3}$ & $\rightarrow$ & $\SO(7)$ \;,\\[1ex]
$\nu=2$ & $\rightarrow$ & $\SO(6)\times\SO(2)$ \;,\\[1ex]
$\nu=4$ & $\rightarrow$ & $\SO(5)\times\SO(3)$ \;,\\[1ex]
$\nu=\infty$ & $\rightarrow$ & $\SO(4)\times\SO(4)$ \;.
\end{tabular}
\end{equation}
The values \,$\nu=\infty$\, or \,$\nu=\pm 2$\, correspond to models which can be embedded in the STU truncation of the ${\rm SO}(8)$ gauged $\mathcal{N}=8$ supergravity.
Therefore the black hole solutions discussed in this work, in the absence of electric and magnetic charges, can all be embedded maximal supergravity. Since the ${\rm SO}(8)$ gauged maximal supergravity can be uplifted to $D=11$ supergravity only for $\omega=0$ \cite{deWit:2013ija}, only for $L\,\alpha=0$ or equivalently for $L\,\alpha=\pm 1$  our solutions can be embedded in the eleven dimensional theory through maximal supergravity, by means of the formulas presented in \cite{Cvetic:1999xx,Cvetic:2000eb}.

\subsection{Charged case}
In this section we show how the charged solutions can be embedded in the dyonic ${\rm SO}(8)$ gauged maximal supergravity.

\paragraph{The $\nu=4$ case.}
Let us focus first on the $\omega=0$ case and eventually discuss the more general one.
The solutions describe gravity coupled to one scalar field and two vector fields. When identified with fields in the maximally supersymmetric model, the scalar and the two vectors should not excite the other fields in the model, such as the scalar fields in the ${\bf 35}_\text{c}$ of ${\rm SO}(8)$.
The latter parametrize ${\rm E}_{7(7)}$ generators which, in the symplectic frame we are working in, are off-diagonal. This means that, in the kinetic Lagrangian of the vector fields, only the $\mathcal{R}_{\Lambda\Sigma}$ matrix has a linear dependence on those scalars, so that the vector fields can source the scalars in the ${\bf 35}_\text{c}$ only through the term $\mathcal{R}_{\Lambda\Sigma}\,F^\Lambda\wedge F^\Sigma$. We thus require, in our solution, the condition:
\begin{equation}
F^\Lambda\wedge F^\Sigma\=0\,,
\label{eq:pseudoscalarcond}
\end{equation}
in order to consistently keep, in the maximal theory, the scalars in the ${\bf 35}_\text{c}$ equal to zero.
Of the remaining scalar fields in the ${\bf 35}_\text{v}$ of ${\rm SO}(8)$, as explained in footnote \ref{so8}, 28 are gauged away so that we are left with the seven independent scalar fields $\vec{\phi}=(\phi_i)$, with $i=1,\dots,8$\, and \,$\sum_{i=1}^8\phi_i=0$, parameterizing the Cartan subalgebra of $\mathfrak{e}_{7(7)}$ (Lie algebra of $E_{7(7)}$).\par
Next we want to further truncate the theory to the scalar field $\phi$ which is singlet with respect to the subgroup ${\rm SO}(5)\times {\rm SO}(3)$ of ${\rm SO}(8)$. In the absence of vector fields this truncation was illustrated in the previous section. When the solution is charged, on the other hand, the two vector fields involved in it should be identified with two of the $28$ vectors on the maximal model which do not source the six scalar fields among $\vec{\phi}$ that we wish to set to zero. Before working this condition out let us observe that any two-dimensional subgroup of ${\rm SO}(8)$ is abelian, so that the two generators $J_1,\,J_2$ gauged by $A^1_\mu$ and $A^2_\mu$ must commute. Another consistency condition on $J_1$ and $J_2$ is that the minimal couplings of $A^1_\mu$ and $A^2_\mu$  to the scalar fields, once these are restricted to $\phi$, must vanish. This amounts to requiring the Killing vectors describing the action of $J_1$ and $J_2$ on the scalars, as isometry generators of the scalar manifold, to be zero. Let $\mathbb{L}(\phi)$ be the coset representative restricted to the scalar $\phi$ only. One can show that the Killing vectors associated with $J_1$ and $J_2$ are zero, once we restrict the scalars to $\phi$, if, and only if, the following condition on the 56-dimensional symplectic representation of $J_\lambda=(J_1,\,J_2)$ and of $\mathbb{L}(\phi)$ holds:
\begin{equation}
\mathbb{L}(\phi)^{-1}\,J_\lambda\:\mathbb{L}(\phi)
   -\mathbb{L}(\phi)\:J_\lambda\:\mathbb{L}(\phi)^{-1}\=\Zero\:,
\end{equation}
which amounts to requiring that $J_\lambda$ ($\lambda=1,2$) commute with the Cartan generator parameterized by $\phi$.\par
Let us now discuss the consistency of the truncation of the dilatonic scalars $\vec{\phi}$ to $\phi$. To this end, let us write $\vec{\phi}$ as follows:
\begin{equation}
\begin{split}
\begin{alignedat}{3}
&\phi_1= -\frac{1}{2}\,\sqrt{\frac{3}{10}}\,\phi +\varphi_1\:,\qquad
&&\phi_2= -\frac{1}{2}\,\sqrt{\frac{3}{10}}\,\phi +\varphi_2\:,\qquad
&&\phi_3=- \frac{1}{2}\,\sqrt{\frac{3}{10}}\,\phi +\varphi_3\:,
\\[\jot]
&\phi_4= -\frac{1}{2}\,\sqrt{\frac{3}{10}}\,\phi +\varphi_4\:,\qquad
&&\phi_5=-\frac{1}{2}\,\sqrt{\frac{3}{10}}\,\phi-\sum_{k=1}^4\varphi_k\:,\qquad
&&\phi_6=\frac{1}{2}\,\sqrt{\frac{5}{6}} \phi+\varphi_5\:,
\\[\jot]
&\phi_7=\frac{1}{2}\,\sqrt{\frac{5}{6}}\phi+\varphi_6\:,\qquad
&&\phi_8=\frac{1}{2}\,\sqrt{\frac{5}{6}} \phi-\varphi_5-\varphi_6\:. &&
\end{alignedat}
\end{split}
\end{equation}
Writing the ${\rm SO}(8)$ generators as $T_{IJ}=-T_{JI}$ ($I,J=1,\dots,8$), the equations for the scalars $\varphi_\ell$ ($\ell=1,\dots 6$) are satisfied when $\varphi_\ell\equiv 0$, and the scalar $\phi$ enters the kinetic terms of the vector fields as in \eqref{eq:canonicact} with $\nu=4$ if $J_1$ and $J_2$ are chosen as follows:
\begin{equation}
J_1=\sqrt{\frac{2}{5}}\,\left(T_{12}+\frac{\varepsilon_1}{\sqrt{2}}\,T_{34}+\frac{\varepsilon_2}{\sqrt{2}}\,T_{35}+\frac{\varepsilon_3}{\sqrt{2}}\,
T_{45}\right)\,,
\qquad\;
J_2=\frac{1}{\sqrt{3}}\,\left(T_{67}+\varepsilon_4\,J_{68}+\varepsilon_5\,J_{78}\right)\,,
\end{equation}
where $\varepsilon_\ell^2=1$. This identifies the two vector fields $A^1_\mu$ and $A^2_\mu$ out of $A^{IJ}_\mu$:
\begin{equation}
\frac{1}{2}\,A^{IJ}_\mu\,T_{IJ}\=A^1_\mu\,J_1+A^2_\mu\,J_2\:,
\end{equation}
so that the two field strengths $\bar{F}^1_{\mu\nu},\,\bar{F}^2_{\mu\nu}$ in (\ref{eq:canonicact}) are identified with the $F^{IJ}_{\mu\nu}$ of the maximal theory as follows:
\begin{equation}
\begin{split}
F^{12}_{\mu\nu}&=\sqrt{\frac{2}{5}}\,\bar{F}^1_{\mu\nu}\,,\quad\;
F^{34}_{\mu\nu}=\frac{\varepsilon_1}{\sqrt{{5}}}\,\bar{F}^1_{\mu\nu}\,,\quad\;
F^{35}_{\mu\nu}=\frac{\varepsilon_2}{\sqrt{{5}}}\,\bar{F}^1_{\mu\nu}\,,\quad\;
F^{45}_{\mu\nu}=\frac{\varepsilon_3}{\sqrt{{5}}}\,\bar{F}^1_{\mu\nu}\,,
\\[2\jot]
F^{67}_{\mu\nu}&=\frac{1}{\sqrt{{3}}}\,\bar{F}^2_{\mu\nu}\,,\quad\;
F^{68}_{\mu\nu}=\frac{\varepsilon_4}{\sqrt{{3}}}\,\bar{F}^2_{\mu\nu}\,,\quad\;
F^{78}_{\mu\nu}=\frac{\varepsilon_5}{\sqrt{{3}}}\,\bar{F}^2_{\mu\nu}\,.
\end{split}
\end{equation}
When $\omega\neq 0$, the same generators $T_{IJ}$ are gauged by linear combinations of $A^{IJ}_\mu$ and $A_{IJ\,\mu}$ of the form
\begin{equation}
\hat{A}_\mu^{IJ}\=\cos(\omega)\,A^{IJ}_\mu-\sin(\omega)\,A_{IJ\,\mu}\:,
\end{equation}
which means that the gauging of ${\rm SO}(8)$ is performed in a different symplectic frame in which the electric vector fields are no longer $A^{IJ}_\mu$ but rather $\hat{A}_\mu^{IJ}$. Let us denote by $\hat{A}^{\hat{M}}_\mu=(\hat{A}^{\hat{\Lambda}}_\mu,\,\hat{A}_{\hat{\Lambda}\,\mu})$, where $\hat{A}^{\hat{\Lambda}}_\mu=\hat{A}_\mu^{IJ}$, the vector fields and their magnetic duals in the new symplectic frame, and by ${A}^{{M}}_\mu=({A}^{{\Lambda}}_\mu,\,{A}_{{\Lambda}\,\mu})$ the same vectors in the old frame. Let $\mathbb{F}^{\hat{M}}_{\mu\nu}=\partial_\mu \hat{A}^{\hat{M}}_\nu-\partial_\nu \hat{A}^{\hat{M}}_\mu$ and $\mathbb{F}^{{M}}_{\mu\nu}=\partial_\mu {A}^{{M}}_\nu-\partial_\nu {A}^{{M}}_\mu$ be the corresponding field strengths. We have the following relation:
\begin{equation}
\mathbb{F}^{{\hat{M}}}_{\mu\nu}\=E_M{}^{{\hat{M}}}\,\mathbb{F}^{{M}}_{\mu\nu}\,,
\end{equation}
where
\begin{equation}
E_M{}^{{\hat{M}}}=\left(\begin{matrix}\cos(\omega)\:\Id_{28\times 28} & \sin(\omega)\:\Id_{28\times 28}\cr-\sin(\omega)\:\Id_{28\times 28}& \cos(\omega)\:\Id_{28\times 28} \end{matrix}\right)\,.
\end{equation}
In the new frame, the parameter $\omega$ will also enter the kinetic matrices $\mathcal{I}_{\hat{\Lambda}\hat{\Sigma}}(\phi,\omega),\,\mathcal{R}_{\hat{\Lambda}\hat{\Sigma}}(\phi,\omega)$. These matrices are indeed defined, through equation (\ref{MRI}), as components of the new symplectic matrix $\mathcal{M}_{\hat{M}\hat{N}}(\phi,\omega)$ which is expressed in terms of the $\omega$-independent $\mathcal{M}_{MN}(\phi)$ in the original frame through the relation:
\begin{equation}
\mathcal{M}_{\hat{M}\hat{N}}(\phi,\omega)\=E^{-1}(\omega)_{\hat{M}}{}^M\,E^{-1}(\omega)_{\hat{N}}{}^N\,\mathcal{M}_{MN}(\phi)\,.
\end{equation}
Upon truncating scalar and vector fields as described above, the two vectors will only enter the bosonic action through the corresponding field strengths. Written in the new symplectic frame, the kinetic terms of $\hat{A}^{\hat{\lambda}}_\mu=(\hat{A}_\mu^1,\,\hat{A}_\mu^2)$ will depend on the $\omega$ parameter through the restrictions \,$\mathcal{I}_{\hat{\lambda}\hat{\sigma}}(\phi,\omega),\,\mathcal{R}_{\hat{\lambda}\hat{\sigma}}(\phi,\omega)$\: of  \:$\mathcal{I}_{\hat{\Lambda}\hat{\Sigma}}(\phi,\omega),\,\mathcal{R}_{\hat{\Lambda}\hat{\Sigma}}(\phi,\omega)$\, to the two vectors.
This dependence can, however, be undone at the level of the bosonic field equations and Bianchi identities since the latter depend on $\mathbb{F}^{\hat{m}}_{\mu\nu}=({F}^{\hat{\lambda}}_{\mu\nu},\,{G}_{\hat{\lambda}\mu\nu})$ only in symplectic-invariant contractions with the matrix $\mathcal{M}_{\hat{m}\hat{n}}(\phi,\omega)$ and its derivatives, see (\ref{eqsofmot}), so that the dependence on $\omega$ of the terms involving the vector field strengths, can be disposed of through a redefinition of the latter which amounts to writing them in terms of the field strengths $\mathbb{F}^m_{\mu\nu}$ in the original frame (consisting of $\bar{F}^1_{\mu\nu},\,\bar{F}^2_{\mu\nu}$ and their magnetic duals), through the matrix $E$.%
\footnote{More precisely though the $4\times 4$ matrix $E_m{}^{{\hat{m}}}$, which has the same form as $E_M{}^{{\hat{M}}}$, though written in terms of $2\times 2$ blocks.}
Upon this redefinition, the bosonic field equations of the truncated model coincide with those obtained from the action (\ref{eq:canonicact}), with $\nu=4$, provided we identify:
\begin{equation}
g=\frac{\sqrt{2}}{L}\:,\qquad\;
\sin(\omega)=L\,\alpha\:,\qquad\;
\cos(\omega)=\sqrt{1-L^2 \alpha^2}\:,
\end{equation}
or, changing $\phi$ into $-\phi$,
\begin{equation}
g=\frac{\sqrt{2}}{L}\:,\qquad\;
\cos(\omega)=L\,\alpha\:,\qquad\;
\sin(\omega)=\sqrt{1-L^2\,\alpha^2}\:.
\label{eq:chargednu4id}
\end{equation}
We still require, in this new symplectic frame, the constraint \eqref{eq:pseudoscalarcond} as a sufficient condition for the consistent truncation of the $\N=8$ pseudoscalar fields in the $\textbf{35}_\text{c\,}$.\par
The embedding of the $\nu=4/3$ model in the maximal theory is more subtle and will be dealt with in a future work.
In the remaining cases \,$\nu=\infty$\, or \,$\nu=\pm2$, our solutions can be indeed extended to charged solutions within $\mathcal{N}=8$, $\SO(8)$ gauged supergravity, within the bosonic part of the STU truncation of it.\par
All the non-supersymmetric solutions presented here have a non-trivial $\omega$-parameter, and we believe they are all new except the $\nu=\pm 2$ and the $\nu=\infty$ case, which were found in the non-dyonic case by Duff and Liu (see Appendix \ref{app:cases}). We have also found that in the supersymmetric limit the $\omega$-parameter is fixed: therefore, it is plausible that these solutions are BPS black holes of the usual de Wit-Nicolai maximal supergravity and can be uplifted to 11 dimensions.
As far as the regular supersymmetric solutions are concerned (Sect.\ \ref{sec:susybh}), we have seen that the regularity condition $\alpha^2=L^{-2}$ can be satisfied by setting $\omega=0$, see eqs.\ \eqref{eq:chargednu4id}.  This implies that those black holes are also solutions to the original De Wit and Nicolai maximal supergravity, whose embedding in the $D=11$ supergravity is well known. This also allows for an eleven-dimensional description of the same solutions.

\section{Conclusions}
In this article we have studied a $\mathcal{N}=2$ prepotential that contains all the single dilaton truncations of $\mathcal{N}=8$ supergravity with a possible dyonic gauging. By generalizing our previous results, we were able to construct electrically and magnetically charged black hole solutions with their corresponding supersymmetric limits. We exploited the non-trivial transformation of the FI parameters under electromagnetic duality to connect the electric and the magnetic solutions. We were able to recover the known results in the case of the STU model, which is the case of $n=1$ and $\nu=\pm 2$. All the other charged solutions are new.  We also constructed a new supersymmetric truncation of the maximal supergravity and found non-extremal and supersymmetric black holes in this new sector. These black holes can be uplifted to higher dimensions and can be interpreted as new M2-brane and D2-brane with a regular supersymmetric limit.
It is important to emphasize that the fact that, for certain models, we did not derive, in our analysis, an embedding in $D=11$ supergravity, does not rule out the existence for them of an UV completion in superstring or M-theory. The definition of such uplift will be object of future investigations.\par
We expect that these infinitely many black holes can be a playground to test ideas of holography, microscopic state counting and condensed matter, as well as a small step towards the characterization of all the supersymmetric solutions of M-theory.

\vspace{1em}
\section*{\normalsize Acknowledgments}
\vspace{-5pt}
We thank Adolfo Guarino for valuable discussions. We would like to thank the support of Proyecto de cooperación internacional 2019/13231-7 FAPESP/ANID. The research of AA is
supported in part by the Fondecyt Grants 1170279 and 1181047 and Banco Santander through Universidad de Oviedo. The research of DA is supported in part by the Fondecyt Grant 1200986.
\vspace{1em}


\appendix
\addtocontents{toc}{\protect\setcounter{tocdepth}{1}}
\addtocontents{toc}{\protect\addvspace{5pt}}
\numberwithin{equation}{section}%
\numberwithin{figure}{section}%

\section{Recovering known cases}\label{app:cases}

\subsection{Duff-Liu} \label{subapp:DL}
One of the first attempts to construct supersymmetric black holes in gauged supergravities was done by Duff and Liu in \cite{Duff:1999gh}. While all their supersymmetric solutions are singular, we found that some of their non-extremal solutions coincide with ours. Let us describe this.\par
The theory considered here is the well known STU model of gauged $\N=8$ supergravity (with no axions), described by the action:
\begin{equation}\label{eq:Action}
\begin{gathered}
\mathscr{S}=\frac{1}{2}\int d^4 x\;\sqrt{-g}\left(-R+\sum\limits_{m=1}^{3}
    \left[\frac{1}{2}\left(\partial\varphi_{m}\right)^{2}
    +\frac{2}{L^{2}}\cosh(\varphi_{m})\right]  -\frac{1}{4}\sum\limits_{i=1}^{4}e^{\vec{a}_{i}\cdot\vec{\varphi}}
    \big(F^{(i)}\big)^{2}\right),
\\[1.5ex]
\vec{\varphi}=(\varphi_{1},\varphi_{2},\varphi_{3}),\quad\;
\vec{a}_{1}=(1,1,1),\quad\;
\vec{a}_{2}=(1,-1,-1),\quad\;
\vec{a}_{3}=(-1,1,-1),\quad\;
\vec{a}_{4}=(-1,-1,1). 
\end{gathered}
\end{equation}
The electric Duff-Liu static black hole solution can be written as follows:
\begin{equation}
\begin{split}
&ds^{2}=\frac{f}{\sqrt{H}}\,dt^{2}-\sqrt{H}\left(\frac{dr^{2}}{f}
    +r^{2}\,d\Sigma_{k}\right)\,,\qquad f=k+\frac{r^{2}}{L^{2}}\,H-\frac{\mu}{r},
\\[\jot]
&H=\,H_1\,H_2\,H_3\,H_4\,,\qquad
H_{i}=1+\frac{S_{i}\,\mu}{r},\qquad
A^{(i)}=\pm\frac{\sqrt{S_{i}+k\,S_{i}^{2}}\:\mu}{r+S_{i}\,\mu}\,dt\,,
\\[2\jot]
&\varphi_{1}=\frac{1}{2}\ln\left(\frac{H_{1}\,H_{2}}{H_{3}\,H_{4}}\right)\,,\qquad
\varphi_{2}=\frac{1}{2}\ln\left(\frac{H_{1}\,H_{3}}{H_{2}\,H_{4}}\right)\,,\qquad
\varphi_{3}=\frac{1}{2}\ln\left(\frac{H_{1}\,H_{4}}{H_{2}\,H_{3}}\right)\,.\qquad
\label{eq:BH4}
\end{split}
\end{equation}

\paragraph{\boldmath The $T^3$ model. \unboldmath}
The well known single scalar field truncation when all scalar fields are equal, the $T^3$ model, is retrieved for $S_2=S_3=S_4$, in which case $\varphi_1=\varphi_2=\varphi_3$. Let us define the scalar field $\phi$
\begin{equation}
\varphi_1=\varphi_2=\varphi_3=-\sqrt{\frac{2}{3}}\,\phi\,,
\end{equation}
and the canonically normalized gauge fields $F^1=\sqrt{2}\,\bar{F}^1$ and $F^2=\sqrt{\frac{2}{3}}\,\bar{F}^2$. Then the STU model reduces to the $T^3$ model
\begin{equation}
\mathscr{S}=\int d^4 x\;\sqrt{-g}  \left(-\frac{R}{2}+\frac{1}{2}\left(\partial\phi\right)^{2}+\frac{3}{L^{2}}\cosh \Big(\sqrt{\frac{2}{3}}\phi\Big) -\frac{1}{4} e^{-3\sqrt{\frac{2}{3}}\phi}\left( \bar{F}^{1}\right)^{2}-\frac{1}{4} e^{\sqrt{\frac{2}{3}}\phi}\left( \bar{F}^{2}\right) ^{2}\right),
\label{eq:Action t3}%
\end{equation}
which exactly coincides with our model for $\nu=-2$. Moreover, our solutions indeed coincide with Duff-Liu in the $\nu=-2$ case, as it can be seen from the change of coordinates
\begin{equation}
x^2=\frac{H_{1}}{H_{2}}\:,
\end{equation}
and the following reparametrization of the integration constants:
\begin{equation}
\eta=\pm \frac{2\,L^2}{\mu(S_1-S_2)}\,,\qquad
Q_1= \pm \sqrt{2}\,\frac{\sqrt{S_1+k\,S_1^2}}{S_1-S_2}\,,\qquad
Q_2 =\pm \sqrt{6}\,\frac{\sqrt{S_2+k\,S_2^2}}{S_1-S_2}\,.
\end{equation}
Duff and Liu found that these solutions are naked singularities in the BPS limit. Their limit is achieved by redefining $S_i=\lambda_i\,\mu^{-1}$ and then taking $\mu\rightarrow0$. This makes the lapse function positive definite for spherical and planar horizons, hence one finds a naked singularity. The ansatz of the Killing spinors of Duff and Liu includes dependence on the coordinates of $d\Sigma_k$. This is the reason of the singular BPS limit, as described in detail in \cite{Hristov:2010ri}, while the Cacciatori-Klemm Killing spinors has only non-trivial radial dependence.

\paragraph{\boldmath The $n=1$\, or \,$\nu=\infty$  model. \unboldmath}
Here we set $S_1=S_3$ and $S_2=S_4$, in which case $\varphi_1=\varphi_3=0$. Our scalar field $\phi$ is
\begin{equation}
\varphi_2=\sqrt{2}\,\phi\,,
\end{equation}
and the canonical normalized gauge fields are $F^1=\bar{F}^1$ and $F^2=\bar{F}^2$. Then the STU model reduces to the $n=1$ model
\begin{equation}
\mathscr{S}=\int d^4 x\;\sqrt{-g}  \left( -\frac{R}{2}+\frac{1}{2}(\partial\phi)^{2} +\frac{1}{L^{2}}\left(2+\cosh(\sqrt{2}\,\phi)\right) -\frac{1}{4}e^{\sqrt{2}\,\phi}\left(\bar{F}^{1}\right)^{2}
-\frac{1}{4}e^{-\sqrt{2}\,\phi}\left(\bar{F}^{2}\right)^{2}\right),
\label{Action n1}
\end{equation}
which indeed coincides with our model for $n=1$. Our solutions coincide with Duff-Liu in this case, provided either the magnetic charge or the electric charge is set to zero. This follows from the change of coordinates
\begin{equation}
x=\frac{H_{2}}{H_{1}}\,,
\end{equation}
with the following reparametrization of the integration constants,
\begin{equation}
\eta=\pm \frac{L^2}{\mu(S_1-S_2)}\,,\qquad\;
Q_1= \pm \frac{\sqrt{S_1+k\,S_1^2}}{S_1-S_2}\,,\qquad\;
Q_2 =\pm \frac{\sqrt{S_2+k\,S_2^2}}{S_1-S_2}\,,
\end{equation}
and setting the magnetic charges to zero. We would like to remark that this connection with Duff and Liu implies that these solutions can also be obtained by adequate truncation of those of Chow and Compere \cite{Chow:2013gba}.

\subsection{Cacciatori-Klemm} \label{subapp:CK}

\paragraph{\boldmath The $T^3$ model. \unboldmath}
The first example of genuine supersymmetric black holes in AdS$_4$ was found by Cacciatori and Klemm \cite{Cacciatori:2009iz}. Let us discuss the supersymmetric solution following the lines of Section 7 of \cite{Hristov:2010ri}, we shall refer to this reference as ``HV'' below.
This supersymmetric solution is spherically symmetric and purely magnetic and can be obtained from our prepotential \eqref{eq:prepotentialn} with $n=\frac{1}{2}$, corresponding to $\nu=-2$, together with $\theta_M=(\xi_0,\xi_1,0,0)$ Fayet-Iliopoulos terms and replacing \,$V\rightarrow g^2\,V$ in the bosonic Lagrangian \eqref{eq:boslagr}.\par
The explicit solution for the dilaton trucation $z=\exp(\lambda\,\phi)$, in terms of canonically normalized fields (using an action with form \eqref{eq:canonicact}), turns out to be
\begingroup
\setlength{\belowdisplayshortskip}{2pt plus 1pt}%
\setlength{\belowdisplayskip}{4pt plus 1pt minus 1pt}%
\begin{equation}
\begin{split}
&\phi(r)\=\frac{1}{\lambda}\,\log\left(\frac{\xi_0}{\xi_1}\,\frac{3\,r\pm4\,\beta_1\,\xi_1}{r\mp4\,\beta_1\,\xi_1}\right)
    +\frac{1}{\lambda}\,\log\left(\frac{\xi_1}{3\,\xi_0}\right)\:,
\\[2ex]
&\bar{F}^1_{\theta\varphi}\,=\frac{1}{2\sqrt{2}}\left(\frac{3\,\xi_0}{\xi_1}\right)^\frac{3}{4}\left(\frac{1}{4\,g\,\xi_0}+\frac{128\,g\,\beta_1^2\,\xi_1^2}{3\,\xi_0}\right)\,\sin(\theta)={\color{Sienna}\frac{1}{2\sqrt{2}}\left(\frac{3\,\xi_0}{\xi_1}\right)^\frac{3}{4}}{\color{LimeGreen!80!black}\sqrt{8}}\;\,2\,p^0\,\sin(\theta)\:,
\\[2ex]
&\bar{F}^2_{\theta\varphi}\,=\frac{1}{2\sqrt{2}}\left(\frac{3\,\xi_1}{\xi_0}\right)^\frac{1}{4}\left(\frac{3}{4\,g\,\xi_1}-\frac{128\,g\,\beta_1^2\,\xi_1}{3}\right)\,\sin(\theta)={\color{Sienna}\frac{1}{2\sqrt{2}}\left(\frac{3\,\xi_1}{\xi_0}\right)^\frac{1}{4}}{\color{LimeGreen!80!black}\sqrt{8}}\;\,2\,p^1\,\sin(\theta)\:,
\\[1.5ex]
&N=\left(\sqrt{8}\,g\,r+\frac{1-\frac{32\,(\sqrt{8}\,g\,\beta_1\,\xi_1)^2}{3}}{2\,\sqrt{8}\,g\,r}\right)^2,
\qquad
H=\frac{(3\,r\pm4\,\beta_1\,\xi_1)^3\, (r\mp4\,\beta_1\,\xi_1)}{4\,\xi_0\,\xi_1^3\,r^4}\:,
\\[1.5ex]
&ds^{2}\=\gamma^2\,\frac{N}{\sqrt{H}}\,dt^{2}-
    \sqrt{H}\,\left(\frac{dr^{2}}{N}+r^2\,d\Sigma_1\right)\:,
\label{eq:solHVD}
\end{split}
\end{equation}
\endgroup
with $\lambda=\sqrt{8/3}$, to make the scalar field canonically normalized, and where we have included the factor $\gamma^2$ in the $g_{tt}$ lapse function, to take into account possibly different time normalizations. Here $p^0$ and $p^1$ are the one given in eq.\ (7.3) of HV. It should be taken into account that the coupling constants are related as $g=g^\textsc{us}=g^\textsc{hv}/\sqrt{8}$. The magnetic charges are multiplied by different factors that we have included with different colours to understand how to match these solutions. The factors in {\color{Sienna}brown} can be traced back to the fact that we use canonically normalized fields, these factors appearing in equations \eqref{cangauge}. The factors in {\color{LimeGreen!80!black}green} comes from the fact that the prepotentials are related as $\mathcal{F}^\textsc{us}=\mathcal{F}^\textsc{hv}/\,8$, which implies that the matrix of the gauge-field couplings are related as \,$\mathcal{I}^\textsc{us}=\mathcal{I}^\textsc{hv}/\,8$. We also note that the Lagrangian of HV has a different normalization of the gauge fields than ours; however, they use the convention $ 2 \: F^\Lambda_{\mu\nu}\=\partial_\mu A^\Lambda_\nu-\partial_\nu A^\Lambda_\mu$,
which cancels this difference when comparing the $\U(1)$ connections. Finally, there is a factor $2$ that we were not able to trace back. We remark that we were able to verify that the configuration \eqref{eq:solHVD} is a solution of the canonically normalized Lagrangian of our paper with
\begin{equation}
L=\frac{ 3^{\frac{3}{4}}}{\sqrt{2}\:\xi_1^{\frac{3}{4}}\:\xi_0^{\frac{1}{4}}g^\textsc{hv}}=\frac{3^{\frac{3}{4}}}{4\: \xi_1^{\frac{3}{4}}\:\xi_0^{\frac{1}{4}}g^\textsc{us}}\:.
\end{equation}
These quantities satisfy the first our BPS conditions \eqref{BPS magnetic}
\begin{equation}
P_1+\sqrt{3}\:P_2-\sqrt{2}\:L\=0\:.
\end{equation}
Let us match this solution with the solution given in the body of our paper. From the equality of the canonically normalized scalar fields we find that
\begin{equation}
-\frac{2}{\lambda}\,\log x \= \frac{1}{\lambda}\,\log\left(\frac{\xi_0}{\xi_1}\,\frac{3\,r\pm4\,\beta_1\,\xi_1}{r\mp4\,\beta_1\,\xi_1}\right)
    +\frac{1}{\lambda}\,\log\left(\frac{\xi_1}{3\,\xi_0}\right)\:,
\end{equation}
which yields $x=x(r)$. From here we find that our $g_{\theta \theta}$ is indeed $r^2\,H(r)$ if
\begin{equation}
\eta^2\=\frac{27\,\sqrt{3}}{2^7\:\xi_0^{\frac{1}{2}}\:\xi_1^{\frac{7}{2}}\:\beta_1^{2}(g^\textsc{hv})^4}=\frac{27\,\sqrt{3}}{2^{13}\:\xi_0^{\frac{1}{2}}\:\xi_1^{\frac{7}{2}}\:\beta_1^{2}(g^\textsc{us})^4}\:.
\end{equation}
Now we can verify our second BPS condition with $\alpha=0$ \eqref{BPS magnetic}:
\begin{equation}
P_2-\frac{\sqrt{3}}{2\sqrt{2}}\:L+\frac{\sqrt{3}}{2\sqrt{2}}\:\frac{L^3}{\eta^2}\=0\:.
\end{equation}
Finally, we have that the metrics are equal provided the product of the lapse function and $g_{\theta \theta}$ is the same in both cases:
\begin{equation}
-g_{tt}\,g_{\theta \theta}\=\gamma^{2}\,N\,r^2\=\Upsilon(x)^2\,f(x)\,L^2\:.
\end{equation}
From our supersymmetric solution it follows that
\begin{equation}
\gamma\,\sqrt{8}\,g\,r^2+\gamma\,\frac{1-\frac{32\,(\sqrt{8}\,g\,\beta_1\,\xi_1)^2}{3}}{2\,\sqrt{8}\,g}
\=\Upsilon(x)\,\frac{x^{1-\nu}}{2\,L\,\nu^2}\,
    \Big((x^\nu-1)^2\,\eta^2+L^{2}\big(2\,x^\nu(\nu^{2}-1)+x^{2\,\nu}(1-\nu)+\nu +1\big)\Big)\,,
\end{equation}
where we took advantage from the fact that $N$ and $f(x)$ are perfect squares. We find both metrics are the same at $\nu=-2$ by fixing
\begin{equation}
\gamma=\frac{3^{\frac{3}{4}}}{\sqrt{2}\:\xi_0^{\frac{1}{4}}\:\xi_1^{\frac{3}{4}}}\:.
\end{equation}

\section{K\"ahler geometry and effective action} \label{app:geom}

\subsection{Special geometry.}\label{subapp:geom}
A special K\"ahler manifold $\Ms_\textsc{sk}$ is the class of target spaces spanned by the complex scalar fields in the vector multiplets of an $\N=2$ four-dimensional supergravity \cite{Strominger:1990pd,Bagger:1983tt,Lauria:2020rhc}.\par
The geometry of $\Ms_\textsc{sk}$ can be described in terms of an \emph{holomorphic section} $\Omega^M(z^i)$ of the characteristic bundle defined over it, which is the product of a symplectic-bundle and a holomorphic line-bundle. The components of $\Omega^M(z^i)$ are written as
\begin{equation}
\Omega^M=
\left(\begin{matrix}
\X^\Lambda \cr \Fbo_\Lambda
\end{matrix}\right)\;, \qqqquad
\Lambda=0,\,\dots,\nv\;,
\end{equation}
while the {K\"ahler potential} and the {K\"ahler metric} have the following general form
\begin{equation} \label{eq:Kg}
\begin{split}
\mathcal{K}(z,\zb)&\:=\,
    -\log\left[\,i\;\overbar{\Omega}^T\,\Cc\;\Omega\,\right]\:=\,
    -\log\left[\,i\,\left(\overbarcal{\X}^\Lambda\,\Fbo_\Lambda-{\X}^\Lambda\,\overbar{\Fbo}_\Lambda\,\right)\right]\;,
\\[1.5\jot]
g_{i\bar{\jmath}}\:&=\:\partial_i \partial_{\bar{\jmath}}\mathcal{K}\;.
\end{split}
\end{equation}
The choice of $\Omega^M(z^i)$ fixes the symplectic frame (i.e.\ the basis of the symplectic fiber space) and, consequently, the non-minimal couplings of the scalars to the vector field strengths in the Lagrangian. In the special coordinate frame, the lower components $\Fbo_\Lambda$ of the section can be expressed as the gradient, with respect to the upper entries $\X^\Lambda$, of a characteristic \emph{prepotential function} $\F(\X^\Lambda)$:
\begin{equation}
\Fbo_\Lambda\=\frac{\partial\F}{\partial \X^\Lambda}\;,
\end{equation}
where the function $\F(\X^\Lambda)$ is required to be homogeneous of degree two. The upper components $\X^\Lambda(z^i)$ are defined modulo multiplication times a holomorphic function and, in this frame, can be used as projective coordinates to describe the manifold: in a local patch in which
$\X^0\neq 0$, we can identify the scalar fields with the ratios $z^i=\X^i/\X^0$.\par\smallskip
A field $\Phi(z,\zb)$ on the K\"ahler manifold is a section of a $\U(1)$-bundle of weight $p$ if it transforms under a K\"ahler transformation as
\begin{equation}
\Phi(z,\zb)\;\;\rightarrow\;\;e^{i\,p\,\Img[f]}\,\Phi(z,\zb)\;,
\end{equation}
and we can define a correspondent $\U(1)$-\emph{covariant derivative} on the bundle as
\begin{equation}
\begin{split}
\D_i \Phi\,\equiv\,\left(\dd_i+\frac{p}{2}\,\dd_i\mathcal{K}\right)\Phi\;,
\qqquad
\D_{\ib}\Phi\,\equiv\,\left(\dd_\ib-\frac{p}{2}\,\dd_\ib\mathcal{K}\right)\Phi\;.
\end{split}
\end{equation}
and introduce a \emph{covariantly holomorphic vector} $\V^M$
\begin{equation}
\V^M\=e^{\frac{\mathcal{K}}{2}}\,\Omega^M\=
    \left(\begin{matrix}
          L^\Lambda \cr M_\Lambda\end{matrix}\right)\;,
\end{equation}
which is section of the $\U(1)$-line bundle with weight $p=1$, satisfying the property:
\begin{equation}
\D_{\ib}\,\V^M\=
    \left(\partial_{\bar{\imath}}-\frac{1}{2}\,\partial_{\bar{\imath}}\mathcal{K}\right)\V^M\=0\;,
\end{equation}
while we also have
\begin{equation}
\D_i\,\V^M \=
     \left(\partial_i+\frac{1}{2}\,\partial_i\mathcal{K}\right)\,\V^M\=
     \left(\begin{matrix}f_i^\Lambda\cr
           h_{i\Lambda}\end{matrix}\right)~\equiv~
    \Ucal_i^M \;,
\end{equation}
$\D_{i},\,\D_{\ib}$ being the above $\U(1)$-covariant derivatives.
Under a K\"ahler transformation defined by a holomorphic function $f(z)$, the section transforms by a corresponding $\U(1)$-transformation:
\begin{equation}
\V^M \;\rightarrow\;\; e^{i\,\Img[f]}\,\V^M\;.
\end{equation}
From its definition and eq.\ \eqref{eq:Kg}, we find that $\V^M$ satisfies the condition
\begin{equation}
\V^T\Cc\,\overbar{\V}\=i\;.
\end{equation}
In particular, the definition of this kind of manifold requires the section $\V^M$ to satisfy also the properties
\begin{equation} \label{eq:VMproperties}
\begin{split}
\D_i\,\Ucal_j&\=i\;\mathcal{C}_{ijk}\,g^{k\bar{k}}\,\overbar{\Ucal}_{\bar{k}}\;,
\\[\jot]
\D_i\,\overbar{\Ucal}_{\jb}&\=g_{i\jb}\,\overbar{\V}\;,
\\[\jot]
\V^T\,\Cc\;\Ucal_i&\=0\;,
\\[\jot]
\Ucal_i^T\,\Cc\;\overbar{\Ucal}_\jb&\:=\,-i\,g_{i\jb}\;,
\end{split}
\end{equation}
where $\mathcal{C}_{ijk}$ is a characteristic covariantly holomorphic tensor with weight $p=2$ which enters the expression of the Riemann tensor and defines the Pauli terms in the Lagrangian involving the gauginos.\par
%
The scalar potential $V(z,\zb)$ reads:
\begin{equation}
V\=\left(g^{i\bar{\jmath}}\,\Ucal_i^M\,\overbar{\Ucal}_{\bar{\jmath}}^N
         -3\,\V^M\,\overbar{\V}^N\right)\theta_M\,\theta_N\,=\,
    -\frac{1}{2}\,\theta_M\,\M^{MN}\,\theta_N-4\,\V^M\,\overbar{\V}^N\theta_M\,\theta_N\;,
\label{eq:Vpotz}
\end{equation}
where $\mathcal{M}^{MN}$, and its inverse $\mathcal{M}_{MN}$, are symplectic, symmetric, negative definite matrices encoding the non-minimal couplings of the scalars $z^i$ to the vector fields in the Lagrangian. In particular $\mathcal{M}_{MN}$ has the following block-structure:
\begin{equation}  \label{M}
\mathcal{M}(\phi)\= (\mathcal{M}(\phi)_{MN}) ~\equiv~
\left(
\begin{matrix}
(\RR\II^{-1}\RR+\II)_{\Lambda\Sigma} &\;\;-(\RR\II^{-1})_\Lambda{}^\Gamma \\ -(\II^{-1}\RR)^\Delta{}_\Sigma & (\II^{-1})^{\Delta\Gamma} \\
\end{matrix}
\right)\;,
\end{equation}
and the matrices $\II,\,\RR$ are those contracting the vector field strengths in \eqref{eq:boslagr}. It is easily verified that the above potential \eqref{eq:Vpotz} can be expressed in terms of a \emph{complex superpotential}
\begin{equation}
\W\=\V^M\,\theta_M\;,
\end{equation}
section of the $\U(1)$-bundle with $p=1$, as follows:
\begin{equation}
V\=g^{i\jb}\,\D_i\W\;\D_{\jb}\overbar{\W}-3\,|\W|^2\;.
\end{equation}
We can also define a \emph{real superpotential} $\mathcalboondox{W}=|\W|$ in terms of which the potential reads:
\begin{equation}
V\=4\,g^{i\bar{\jmath}}\,\partial_i\mathcalboondox{W}\,\partial_{\bar{\jmath}}\mathcalboondox{W}
    -3\,\mathcalboondox{W}^2\;.
\label{eq:realsuper}
\end{equation}
The introduced $\theta_M$ terms transform in a symplectic representation $\Rsvst$ of the isometry group $G_\textsc{sk}$ of $\Ms_\textsc{sk}$ on contravariant vectors. These FI terms are analogous to the electric and magnetic charges, but while the latter can be considered as solitonic charges of the solution, the former are background quantities actually entering the Lagrangian. Moreover, even though they couple the fermion fields to the vectors, the FI terms do not define vector-scalar minimal couplings.

\subsection{Effective action.}\label{subapp:act}
Let us consider static dyonic black hole configurations and assume a radial dependence for the scalar fields, $z^i=z^i(r)$. The most general metric ansatz, with spherical or hyperbolic symmetry, has the form
\begin{equation}
ds^2\=e^{2\,U(r)}\,dt^2-e^{-2\,U(r)}\left(dr^2+e^{2\,\Psi(r)}\,d\Sigma_k^2\right)\;,
\label{eq:metransatz}
\end{equation}
where \;$d\Sigma_k^2=d\vartheta^2+f_k^2(\vartheta)\,d\varphi^2$\; is the metric on the $2D$-surfaces $\Sigma_k=\{\mathbb{S}^2,\,\mathbb{H}^2,\,\mathbb{R}^2\}$ (sphere, Lobachevskian plane and flat space) with constant scalar curvature \,$R=2\,k$\, and
\begin{equation}
f_k(\vartheta)\=\frac{1}{\sqrt{k}}\,\sin(\sqrt{k}\,\vartheta)
\;\;\rightarrow\;\;
\left\{
\begin{aligned}
\;&\sin(\vartheta)\;,\quad &k&=+1\;;
\\
\;&\sinh(\vartheta)\;,\quad &k&=-1\;.
\\
\;&\;\vartheta\;,\quad &k&=0\;;
\end{aligned}
\right.
\end{equation}
The Maxwell equations are satisfied using the following expression for $\FF^M$
\begin{equation}
\FF^M=\left(\begin{matrix}F^\Lambda\cr G_\Lambda\end{matrix}\right)
    =\:e^{2(U-\Psi)}\,\Cc^{MP}\,\M_{PN}\,\Gamma^N\;dt\wedge dr
      +\Gamma^M\,f_k(\vartheta)\;d\vartheta\wedge d\varphi
      \=d\mathbb{A}^M\;.
\label{eq:FFM}
\end{equation}
The electric and magnetic charges can be defined as
\begin{equation}
e_\Lambda\equiv\frac{1}{\text{vol}(\Sigma_k)}\,\int_{\Sigma_k}G_{\Lambda}\;,
\qqquad
m^\Lambda\equiv\frac{1}{\text{vol}(\Sigma_k)}\,\int_{\Sigma_k}F^{\Lambda}\;,
\end{equation}
where \,$\text{vol}(\Sigma_k)={\displaystyle \int} f_k(\vartheta)\,d\vartheta\wedge d\varphi$.\; The charges can be arranged in the symplectic vector
\begingroup
\belowdisplayskip=12pt
\belowdisplayshortskip=12pt
\begin{equation}
\Gamma^M\=
\left(\begin{matrix}
m^\Lambda \cr
e_\Lambda
\end{matrix}\right)
\=\frac{1}{\text{vol}(\Sigma_k)}\,\int_{\Sigma_k} \FF^M \;.
\label{eq:charges}
\end{equation}
\endgroup
We can obtain the equations of motion coming from the bosonic gauged Lagrangian \eqref{eq:boslagr}, with the metric ansatz \eqref{eq:metransatz}, from a one-dimensional effective action that, apart from total derivative terms, has the form
\begin{equation}
\mathscr{S}_\text{eff}\=
    \mathlarger{\int} dr\,\Lagr_\text{eff}\=
    \mathlarger{\int} dr
    \,\bigg[\,e^{2\,\Psi}\,
        \Big(U^{\prime 2}-\Psi^{\prime 2}
        +g_{i\bar{\jmath}}\,z^{\prime  i}\,\zb^{\prime \,\bar{\jmath}}\,
        \Big)
        -V_\text{eff}\,\bigg]\;,
\label{eq:effact}
\end{equation}
where the prime stands for derivative w.r.t.\ $r$ and where we can define an effective potential
\begin{equation}
V_\text{eff}\,=\,-\,e^{2(U-\Psi)}\,\VBH\,-\,e^{-2(U-\Psi)}\,V\,+\,k\;,
\end{equation}
in terms of the scalar potential $V$ and the (charge-dependent) black hole potential $\VBH$. The latter can be written in the symplectically covariant form
\begin{equation}
\VBH\:=\,-\,\frac{1}{2}\,\Gamma^T\M(\phi)\;\Gamma\;,
\label{eq:VBH}
\end{equation}
in terms of the magnetic and electric charges and scalar-dependent matrix $\M(\phi)$.\par
Once given the effective action, one can make use of the Hamilton-Jacobi formalism and derive a system of first-order equations (\emph{flow equations}) for the warp factors $U(r)$, $\Psi(r)$ and scalar fields $z^i(r)$, $\zb^{\bar{\jmath}}(r)$.
\par

\section{Supersymmetry relations}\label{app:Susycond}
When interested in analysing supersymmetric configurations, one has to impose the vanishing of the SUSY variations, that are written as:
\begin{subequations}\label{eq:susyvar}
\begin{align}
\delta\psi_{\mu A}&\=
        D_\mu\eps_A\+i\;T^{-}_{\mu\nu}\;\gamma^\nu\,\veps_{AB}\,\eps^B
        \+i\;\mathbb{S}_{AB}\,\gamma_\mu\,\eps^B\;,
\\[2ex]
\delta\lambda^{iA}&\=
        i\,\partial_\mu z^i\,\gamma^\mu\,\eps^A \-\frac{1}{2}\;g^{i\jb}\;{\bar{f}}^\Lambda_\jb\;\II_{\Lambda\Sigma}\;F^{-\Sigma}_{\mu\nu}\,\gamma^{\mu\nu}\,\veps^{AB}\,\eps_B
        \+W^{iAB}\,\eps_B\;,
\end{align}
\end{subequations}%
%
%
where we have considered properties \eqref{eq:VMproperties}. The covariant derivatives are written as
\begin{equation}
D_\mu\eps_A\=\dm\eps_A+\frac14\,\spc\,\gamma_{ab}\,\eps_A
    +\frac{i}{2}\,\left(\sigma^2\right)_A{\!}^B\,\mathbb{A}_\mu^M\,\theta_M\,\eps_B+\frac{i}{2}\,\mathcal{Q}_\mu\,\eps_A\;,
\end{equation}
with
\begin{equation}
\mathcal{Q}_\mu\=\frac{i}{2}\left(\dd_{\ib}\mathcal{K}\,\dm\zb^\ib-\dd_{i}\mathcal{K}\,\dm{z}^i\right)\;,
\end{equation}
and, in the chosen parametrization, we also have
%
\begin{equation}
\begin{split}
F^{\pm}_{\mu\nu}&\=\frac12\left(F_{\mu\nu}\pm\Hodge{\!F}_{\mu\nu}\right)\;,  \qquad
    \gamma^{\mu\nu}\=\gamma^{[\mu}\gamma^{\nu]}\;,
\\[2ex]
T_{\mu\nu}&\=L^\Lambda\;\II_{\Lambda\Sigma}\;F^{\Sigma}_{\mu\nu}
    \=\frac{1}{2i}\,L^\Lambda\left(\mathfrak{N}-\overbar{\mathfrak{N}}\right)_{\Lambda\Sigma}\,F^{\Sigma}_{\mu\nu}
    \:=\,-\frac{i}{2}\left(M_\Sigma\,F^{\Sigma}_{\mu\nu}-L^\Lambda\,G_{\Lambda\mu\nu}\right)
    \=\frac{i}{2}\,\V^M\;\Cc_{MN}\;\FF^{N}_{\mu\nu}\;,
\\[2ex]
T^{-}_{\mu\nu}&\=L^\Lambda\;\II_{\Lambda\Sigma}\;F^{-\Sigma}_{\mu\nu}\=\frac{i}{2}\,\V^M\;\Cc_{MN}\;\FF^{-N}_{\mu\nu}\;,
\\[2ex]
T_{i\,\mu\nu}&\=\D_i T_{\mu\nu}
    \=f^\Lambda_i\;\II_{\Lambda\Sigma}\;F^{\Sigma}_{\mu\nu}
    \:=\,-\frac{i}{2}\left(h_{i\Sigma}\,F^{\Sigma}_{\mu\nu}-f^\Lambda_i\,G_{\Lambda\mu\nu}\right)
    \=\frac{i}{2}\;\Ucal_i^M\;\Cc_{MN}\;\FF^{N}_{\mu\nu}\;,
\\[2ex]
\mathbb{S}_{AB}&\=\frac{i}{2}\,\left(\sigma^2\right)_A{\!}^C\;\veps_{BC}\;\theta_M\,\V^M
    \=\frac{i}{2}\,\left(\sigma^2\right)_A{\!}^C\;\veps_{BC}\;\W\;,
\\[2.5ex]
W^{i\,AB}&\=i\,\left(\sigma^2\right)_C{\!}^B\;\veps^{CA}\;\theta_M\;g^{i\jb}\;\overbar{\Ucal}^M_\jb\;,
\end{split}
\end{equation}
%
having used properties
\begin{equation}
\overbar{\mathfrak{N}}_{\Lambda\Sigma}\,F^{-\Sigma}=\,G^{-}_\Lambda\;, \qqquad
L^\Lambda\,\mathfrak{N}_{\Lambda\Sigma}\,=\,M_\Sigma\;.
\end{equation}
The kinetic matrix \,$\mathfrak{N}=\RR+i\,\II$\, can be expressed as \cite{Gaillard:1981rj}
\begin{equation}
\mathfrak{N}_{\Lambda\Sigma}\=
    \dd_{\bar{\Lambda}}\dd_{\bar{\Sigma}}\overbarcal{\F}
    +2\,i\;\frac{\Img\left[\dd_{\Lambda}\dd_{\Gamma}\F\right]\;\Img\left[\dd_{\Sigma}\dd_{\Delta}\F\right]\;L^\Gamma\,L^\Delta}{\Img\left[\dd_{\Delta}\dd_{\Gamma}\F\right]\;L^\Delta\,L^\Gamma}\;,
\end{equation}
with \,$\dd_{\Lambda}=\dfrac{\dd}{\dd\X^\Lambda}$\,, \,$\dd_{\bar{\Lambda}}=\dfrac{\dd}{\dd\bar{\X}^\Lambda}$\,.
\par\smallskip
Just as we did for electric-magnetic charges in \eqref{eq:charges},
we define the \emph{central} and \emph{matter charges}:
\begin{equation}
\begin{split}
\mathscr{Z}&\=\frac{1}{\text{vol}(\Sigma_k)}\,\int_{\Sigma_k}\!\!T
    \=\V^M\,\Cc_{MN}\,\Gamma^N
    \=L^\Lambda\,e_\Lambda - M_\Lambda\,q^\Lambda\;,
\\[1em]
\mathscr{Z}_i&\=\frac{1}{\text{vol}(\Sigma_k)}\,\int_{\Sigma_k}\!\!T_i
    \=\D_{i}\mathscr{Z}
    \=f_i^\Lambda\,e_\Lambda - h_{\Lambda i}\,q^\Lambda\;.
\end{split}
\end{equation}
These are composite quantities that can be thought of as the physical charges measured on a solution at radial infinity. The black hole potential \eqref{eq:VBH} can be schematically rewritten in terms of the central charges as \cite{Ferrara:1997tw,Andrianopoli:2006ub}
\begin{equation}
\VBH\=\left|\D\Zch\right|-|\Zch|^2\;.
\end{equation}
From an explicit computation of the supersymmetry variations \eqref{eq:susyvar}, we find the following relations for the warp factors \cite{Gallerati:2019mzs}
\begin{equation}
\begin{split}
U'&\=e^{U-2\Psi}\;\Real\left[e^{-i\hat{\alpha}}\,\Zch\right]
    +e^{-U}\;\Img\left[e^{-i\hat{\alpha}}\,\W\right]\;,
\\[1.5ex]
\Psi'&\=2\,e^{-U}\,\Img\left[e^{-i\hat{\alpha}}\,\W\right]\;,
\end{split}
\end{equation}
and for the scalars
\begin{equation}
z^{\prime \,i}
    \=e^{-U}\,e^{i\hat{\alpha}}\,g^{i\jb}\;\D_\jb\left(e^{2U-2\Psi}\,\overbarcal{\Zch}-i\,\overbar{\W}\right)\;,
\end{equation}
the above covariant derivative acting on objects with weight $p=-1$, and having introduced two projectors relating the spinor components as \cite{Romans:1991nq,Hristov:2010eu,Hristov:2010ri}
\begin{equation}
\begin{split}
\gamma^0\,\eps_A&\=i\,e^{i\hat{\alpha}}\,\veps_{AB}\,\eps^B\;,
\\[1ex]
\gamma^1\,\eps_A&\=e^{i\hat{\alpha}}\,\delta_{AB}\,\eps^B\;,
\end{split}
\end{equation}
where $\hat{\alpha}$ is an arbitrary constant phase.
The Killing spinors must satisfy the relations \cite{Gallerati:2019mzs}
\begingroup
\belowdisplayskip=4pt
\belowdisplayshortskip=4pt%
\begin{equation}
\begin{split}
\eps_A&\=\chi_A\;e^{\,\frac12\left(U-i\ml{\int}dr\,\mathcal{B}\right)}\;,
\\[1ex]
\eps^A&\=i\,e^{-i\hat{\alpha}}\,\veps^{AB}\,\gamma^0\,\eps_B\;,
\end{split}
\end{equation}
\endgroup
where we have
\begingroup
\abovedisplayskip=4pt
\abovedisplayshortskip=4pt
\begin{equation}
\begin{split}
\dd_r\chi_A&\=0\;,
\\[1ex]
\mathcal{B}&\=\mathcal{Q}_r+2\,e^{-U}\,\Real\left[e^{-i\hat{\alpha}}\,\W\right]\;,
\end{split}
\end{equation}
\endgroup
and the following expression for the phase $\hat{\alpha}$ holds:
\begin{equation}
\dd_r\hat{\alpha}\:=\,-\mathcal{B}\;.
\end{equation}
From the SUSY variations we obtain the property \cite{Gallerati:2019mzs}
\begin{equation}
\Img\left[e^{-i\hat{\alpha}}\,\Zch\right]\:=\,
    -e^{2\Psi-2U}\Real\left[e^{-i\hat{\alpha}}\,\W\right]\;,
\end{equation}
and using also ansatz \eqref{eq:FFM} for $\FF^M$, we find for the $\mathbb{A}^M_\mu$ components:
\begin{equation}
\begin{split}
\mathbb{A}^M_t\,\theta_M&\=2\,e^U\,\Real\left[e^{-i\hat{\alpha}}\,\W\right]\;,
\\[2ex]
\mathbb{A}^M_r&\=0\;,
\\[2ex]
\mathbb{A}^M_\vartheta&\=0\;,
\\[1ex]
\mathbb{A}^M_\varphi&\:=\,-\frac{\Gamma^M}{k}\,\cos\left(\sqrt{k}\,\vartheta\right)\;,
\end{split}
\end{equation}
together with the relation
\begin{equation}
\Gamma^M\,\theta_M\=k\;.
\end{equation}


\hypersetup{linkcolor=blue}
\phantomsection 
\addcontentsline{toc}{section}{References}
\bibliographystyle{mybibstyle}
\bibliography{bibliografia}

\end{document}